\documentclass[a4paper,fleqn,usenatbib]{mnras}
\usepackage[T1]{fontenc}
\usepackage{ae,aecompl}

\usepackage{graphicx}	% Including figure files
\usepackage{amsmath}	% Advanced maths commands
\usepackage{amssymb}	% Extra maths symbols

\usepackage{color}

\newcommand{\sub }[1]{_{\rm{#1}}}

\newcommand{\Msun}{M_\odot}

\newcommand{\Mstar}{M_{\ast}}

\newcommand{\Mref}{\left(\frac{\Mstar}{\Msun}\right)}

\newcommand{\tenau}{\left(\frac{r}{10\,{\rm au}}\right)}
\newcommand{\Mgdot}{\dot{M}_{\rm gas}}
\newcommand{\Mgdotref}{\left(\frac{\Mgdot}{10^{-6}\,\Msun\,{\rm yr^{-1}}}\right)}
\newcommand{\Mddot}{\dot{M}\sub{dust}}
\newcommand{\Mddotref}{\left(\frac{\Mddot}{10^{-4}M_{\oplus}\,{\rm yr^{-1}}}\right)}
\newcommand{\rhoint}{\rho_{\rm int}}
\newcommand{\rhoref}{\left(\frac{\rhoint}{0.1\,\gcmth}\right)}
\newcommand{\adust}{a\sub{dust}}
\newcommand{\Hgas}{\hat{H}\sub{gas}}
\newcommand{\Hdust}{\hat{H}\sub{dust}}
\newcommand{\cs}{c\sub{s}}
\newcommand{\kB}{k\sub{B}}
\newcommand{\MEarth}{M_\oplus}
\newcommand{\fsolar}{f\sub{solar}}
\newcommand{\Minit}{M\sub{init}}
\newcommand{\vrdust}{v_{r,{\rm dust}}}

\newcommand{\St}{{\rm St}}
\newcommand{\mfp}{\lambda\sub{mfp}}
\newcommand{\taccth}{t\sub{acc,3D}}
\newcommand{\tacctw}{t\sub{acc,2D}}

\newcommand{\gcmth}{\rm g\,cm^{-3}}
\newcommand{\gcmtw}{\rm g\,cm^{-2}}
\newcommand{\cms}{\rm cm\,s^{-1}}
\newcommand{\Msunyr}{\Msun\,{\rm yr^{-1}}}

\title[Pebble accretion in class 0/I YSOs]{Pebble accretion in class 0/I YSOs as a possible pathway for early planet formation}
\author[Tanaka \& Tsukamoto]{
Yuki A. Tanaka,$^{1,2}$\thanks{E-mail: tanaka.y.cn@m.titech.ac.jp}
and Yusuke Tsukamoto$^{2}$
\\
$^{1}$Department of Earth and Planetary Sciences, Tokyo Institute of Technology, Meguro, Tokyo, 152-8551, Japan\\
$^{2}$Graduate School of Science and Engineering, Kagoshima University, 1-21-40, Korimoto,  Kagoshima, 890-0065, Japan}

\date{Accepted XXX. Received YYY; in original form ZZZ}

\pubyear{2019}

\begin{document}
\label{firstpage}
\pagerange{\pageref{firstpage}--\pageref{lastpage}}
\maketitle

\begin{abstract}
Recent theoretical works suggest that the pebble accretion process is important for planet formation in protoplanetary disks, because it accelerates the growth of planetary cores.
While several observations reveal axisymmetric sharp gaps in very young disks, which may be indicative of the existence of planets.
We investigate the possibility of planet formation via pebble accretion in much earlier phases, the gravitationally unstable disks of class 0/I young stellar objects.
We find that under the conditions of the class 0/I disks, the pebble accretion timescales can be shorter compared to the typical protoplanetary disks due to larger gas and dust accretion rate, but also find that the accretion timescale is not always a decreasing function of the gas accretion rate.
By using estimated accretion timescales, we give a required initial mass to form cores of gas giants within the lifetime of class 0/I phases under several parameters, such as radial distances from the host star, gas accretion rates, and dust-to-gas mass ratio.
In the most optimistic case, for example the dust-to-gas mass ratio is $f=3\fsolar$, $\sim10^{-4}\MEarth$ objects at 10~au can grow to $10\MEarth$ cores during the typical lifetime of the class 0/I phases, 0.5 Myr.
\end{abstract}

\begin{keywords}
Planetary systems -- planets and satellites: formation -- planets and satellites: physical evolution -- planet-disc interactions
\end{keywords}

%%%%%%%%%%%%%%%%%%%%%%%%%%%%%%%%%%%%%%%%%%%%%%%%%%

%%%%%%%%%%%%%%%%% BODY OF PAPER %%%%%%%%%%%%%%%%%%

%%%%%%%%%%%%%%%%%%%%%%%%%%%%%%%%%%%%%%%%%%%%%%%
%%%%%%%%%%%%%%%%%%%%%%%%%%%%%%%%%%%%%%%%%%%%%%%
%%%%%%%%%%%%%%%%%%%%%%%%%%%%%%%%%%%%%%%%%%%%%%%

\section{Introduction}\label{introduction}

%%%%%%%%%%%%%%%%%%%%%%%%%%%%%%%%%%%%%%%%%%%%%%%
%%%%%%%%%%%%%%%%%%%%%%%%%%%%%%%%%%%%%%%%%%%%%%%
%%%%%%%%%%%%%%%%%%%%%%%%%%%%%%%%%%%%%%%%%%%%%%%

Planets are formed in a protoplanetary disk.
In the traditional planet formation models, it is assumed that kilometer-sized planetesimals are formed from dust particles, after which they grow up to become planetary embryos and/or protoplanets \citep[e.g.,][]{saf72,hay85}.
In particular, in the formation of giant planets, planetary cores form and grow via runaway and oligarchic growth phases \citep[e.g.,][]{ws89,ki98}, and subsequent disk gas accretion onto the cores forms gas giant planets \citep[e.g.,][]{miz80,ste82,iko00}.
This process is the so-called core accretion scenario.
The formation processes of planetary systems are being actively studied both theoretically and observationally, but there still remain several issues to be resolved.
For example, the timescales to form the giant planets by core accretion become too long in the distant orbit in the gas-free condition \citep{dod09,raf11}.

Recently, an alternative model has been proposed; small dust particles directly form large clumps via streaming instability \citep{yg05,joh07,joh12,joh15,car17}, and the clumps grow fast by the accretion of migrating centimeter-sized pebbles that is enhanced by aerodynamic effects \citep{ok10,lj12,lj14,lam14}.
The latter process is the so-called pebble accretion.
In this process, the timescale of the growth of the cores of giant planets is shortened even in the distant region of the disks \citep{lj12}.
In addition, the pebble accretion model has been applied to other conditions, such as the formation of super-Earths systems \citep[e.g.,][]{ct14} and water delivery to the Earth and/or terrestrial embryos \citep{mor16,sat16}.

Planet formation via pebble accretion and its applications in protoplanetary disks (here we refer to protoplanetary disks as disks in the class II/III young stellar objects (YSOs), which is characterized by the absence of mass supply from the envelope.) have been studied \citep[e.g.,][]{gui14,ida16}.
However, the possibility of pebble accretion in class 0/I YSOs, the earlier phases of star and planet formation, has not been considered so far.
It is shown that planetesimal formation from porous dust aggregate \citep{oku12,kat13} is more preferable in the class 0/I phase \citep{tsu17}.
This motivates us to investigate the possibility of pebble accretion in the class 0/I phases for the first time.

Another and the most important motivation of this work is the recent remarkable discoveries of disks with multiple rings and gaps, mainly achieved by the Atacama Large Millimeter/submillimeter Array (ALMA) \citep[e.g.,][]{alm15,and16,ise16,loo17,cie17,fed18}.
One of the most intriguing objects among these is HL Tau, the circumstellar disk whose several axisymmetric sharp gaps and rings have been imaged \citep{alm15}.
In addition, the HCO$^{+}$ observation obtained by ALMA also suggests the existence of a gap structure in the gas component \citep{yen16}.
Many mechanisms have been suggested to reproduce the gap and ring structures found around HL Tau and other objects; for example, magnetorotational instability \citep{joh09,uri11}, secular gravitational instability \citep{you11,ti14,ti16,tom18}, baroclinic instability \citep{kl07,dzy10,flo15}, dust sintering \citep{oku16}, and photoevaporation \citep{erc17}.
Another possible gap-carving mechanism is embedded planets in the protoplanetary disks \citep[e.g.,][]{gt80,lp86}.
Recent works suggest that if gaseous planets exist in the disks, clear gaps similar to those seen in the HL Tau disk and other disks can be opened by the planets \citep[e.g.,][]{dip15,kan15,don15,pk15,jin16,kan16}.

However, the radii of the observed gaps are typically several tens of astronomical units (au), and whether planet formation at such large distances within the early phase of disk evolution is possible or not is still unclear.
This is because the timescale to form the giant planets by classical core accretion scenario, which can be applied for the gas-free planetesimal accretion, is significantly long \citep{dod09,raf11}.
It is suggested that the formation of giant planets via gravitational instability instead of core accretion can account for gaseous planets in distant orbits \citep{sw08,sw09,vb10,tsu13,tsu15}.
For example, planet formation by gravitational instability might take place in $\gtrsim$50 au, and inward migration of the planets can explain the existence of the planets responsible for the gaps \citep{aki16}.
However, the mass of the planet formed via fragments of gravitational instability is typically several Jupiter mass and is too large to explain the properties of the gaps in the HL Tau disk.
For example, \citet{kra10} pointed out that fragments of gravitational instability are unlikely to produce planetary mass objects.

In this work we investigate formation of solid cores go gaseous planets by pebble accretion in class 0/I disks.
It is estimated that the ages of several disks that have clear gaps are very young, such as HL Tau \citep[$\lesssim$1 Myr, e.g.,][]{alm15}, AS 209 \citep[$\sim$0.5 -- 1 Myr,][]{fed18}, Elias 2-24 \citep[$\sim$0.4 Myr,][]{cie17,and10,sie00}.
If planet formation in the class 0/I phases is possible, the existence of the gaps around the class I YSOs can be explained by the giant planet.

This paper is organized as follows.
In Section \ref{diskmodel}, we apply pebble accretion theory to the class 0/I YSOs disk.
Then we investigate the accretion timescales and give the required initial masses to form the cores of the giant planets within the lifetime of the disks, and show their parameter dependence in Section \ref{results}.
Section \ref{discussion} is devoted to a discussion on comparison to the class II disks, migration timescales, the possibility of self-filtration effects, and the possibility of planet formation in the class 0/I phases.
Finally, we summarize our work in Section \ref{conclusion}.

%%%%%%%%%%%%%%%%%%%%%%%%%%%%%%%%%%%%%%%%%%%%%%%
%%%%%%%%%%%%%%%%%%%%%%%%%%%%%%%%%%%%%%%%%%%%%%%
%%%%%%%%%%%%%%%%%%%%%%%%%%%%%%%%%%%%%%%%%%%%%%%

\section{Models}\label{diskmodel}

%%%%%%%%%%%%%%%%%%%%%%%%%%%%%%%%%%%%%%%%%%%%%%%
%%%%%%%%%%%%%%%%%%%%%%%%%%%%%%%%%%%%%%%%%%%%%%%
%%%%%%%%%%%%%%%%%%%%%%%%%%%%%%%%%%%%%%%%%%%%%%%

In this section, we describe our theoretical framework.
For the gas and dust disk model, we assume steady state structures for the gravitationally unstable disk and the dust disk in which radial migration and dust growth are taken into account, respectively \citep{tsu17}.
Theoretically, it has been pointed out that the gravitationally unstable disk frequently forms in class 0/I YSOs \citep[e.g.,][]{nn94,vor09,vb10}
Therefore, we adopt the gravitationally unstable disk.
Some reader may question this point because of the disk mass estimate from observation \citep[e.g.,][]{ans16}.
However, most of the disk mass is estimated from dust thermal emission and is highly uncertain.
We believe that the disk mass used in this study is not unrealistic (see \citet{tsu17} for more detail).

%%%%%%%%%%%%%%%%%%%%%%%%%%%%%%%%%%%%%%%%%%%%%%%
%%%%%%%%%%%%%%%%%%%%%%%%%%%%%%%%%%%%%%%%%%%%%%%

\subsection{Gas disk model}\label{diskmodel2}

%%%%%%%%%%%%%%%%%%%%%%%%%%%%%%%%%%%%%%%%%%%%%%%
%%%%%%%%%%%%%%%%%%%%%%%%%%%%%%%%%%%%%%%%%%%%%%%

%%%%%%%%%%%%%%%%%%%%%%%%%%%%%%%%%%%%%%%%%%%%%%%

\subsubsection{Temperature profile}\label{disktemp}

%%%%%%%%%%%%%%%%%%%%%%%%%%%%%%%%%%%%%%%%%%%%%%%

First, we describe the temperature profile of the disk in our model.
Here, we assume that the disk temperature $T$ is determined by the stellar irradiation.
We assume the temperature profile of the irradiated disk is given as follows:
\begin{eqnarray}
T=T\sub{irr}&\simeq&150\left(\frac{r}{1\,{\rm au}}\right)^{-3/7}\,{\rm K}\nonumber\\
&\simeq&56\tenau^{-3/7}\,{\rm K}
\label{eq:temp}
\end{eqnarray}
This temperature profile is given by \citet{kus70} and \citet{cg97}.

The speed of sound $\cs$ is assumed to be
\begin{eqnarray}
\cs=\sqrt{\frac{\kB T}{m\sub{g}}}\simeq1.9\times10^{4}\left(\frac{T}{10\,{\rm K}}\right)^{1/2}\,\cms,
\label{eq:cs0}
\end{eqnarray}
where $\kB$ is the Boltzmann constant, and $m\sub{g}$ ($\sim\!3.9\times10^{-24}$ g) is the mean molecular mass of the gas.
Substituting equation (\ref{eq:temp}) into (\ref{eq:cs0}), the radial dependence of $\cs$ is given as follows:
\begin{eqnarray}
\cs\simeq4.5\times10^{4}\tenau^{-3/14}\,\cms
\label{eq:cs}
\end{eqnarray}

The aspect ratio of the gas disk $\Hgas$ is as follows:
\begin{eqnarray}
\Hgas\equiv\frac{H\sub{gas}}{r}=\frac{\cs}{r\Omega}
\simeq4.8\times10^{-2}\Mref^{-1/2}\tenau^{2/7},
\label{eq:aspectratio}
\end{eqnarray}
where $H\sub{gas}$, $\Omega$, $\Mstar$, and $\Msun$ are the pressure scale height of the gas disk, rotational frequency, stellar mass, and solar mass, respectively.
The scale height of the gas disk is defined as $\rho\sub{gas}\!\left(z\right)\propto \exp\!\left(-z^{2}/2H\sub{gas}^{2}\right)$.
Here, we assume the Keplerian rotation $\Omega\sub{K}$ for $\Omega$:
\begin{eqnarray}
\Omega\sub{K}&=&\sqrt{\frac{G\Mstar}{r^{3}}}\nonumber\\
&=&2.0\times10^{-1}\Mref^{1/2}\tenau^{-3/2}\,{\rm yr^{-1}},
\label{eq:Omega}
\end{eqnarray}
where $G$ is the gravitational constant.

%%%%%%%%%%%%%%%%%%%%%%%%%%%%%%%%%%%%%%%%%%%%%%%

\subsubsection{Gas disk structure}\label{gasdisk}

%%%%%%%%%%%%%%%%%%%%%%%%%%%%%%%%%%%%%%%%%%%%%%%

We use a steady-state solution for the marginally gravitationally unstable gas disk given by \citet{tsu17} in which $Q=2$ is assumed in entire disks, where $Q$ is Toomre's $Q$ value which is given by
\begin{eqnarray}
Q\equiv\frac{\cs\kappa\sub{ep}}{\pi G\Sigma\sub{gas}},
\label{eq:Q}
\end{eqnarray}
where $\kappa\sub{ep}$ is the epicycle frequency, and $\Sigma\sub{gas}$ is the surface density of the gas.
In this work, we use the approximation $\kappa\sub{ep}=\Omega$.
Assuming that the disk can be described by the viscous $\alpha$ accretion disk model \citep{ss73} and that the gas accretion rate $\Mgdot$ is constant, the following equation should be satisfied:
\begin{eqnarray}
\left|\frac{d\ln\Omega}{d\ln r}\right|\alpha\frac{\cs^{2}}{\Omega}\Sigma\sub{gas}=\frac{1}{2\pi}\Mgdot={\rm const.}
\label{eq:steady}
\end{eqnarray}
From equation (\ref{eq:Q}) with the assumption $Q=2$, we obtain the radial dependence of the gas surface density as follows:
\begin{eqnarray}
\Sigma\sub{gas}\simeq6.7\times10^{2}\Mref^{1/2}\tenau^{-12/7}\,\gcmtw,
\label{eq:sigmag}
\end{eqnarray}
and from equations (\ref{eq:Omega}) and (\ref{eq:steady}), $\alpha$ can be described as
\begin{eqnarray}
\alpha\simeq3.2\times10^{-2}\Mgdotref\tenau^{9/14}.
\label{eq:alpha}
\end{eqnarray}

Because the gas component experiences both the gravity of the central star and the pressure gradient force, the azimuthal velocities between the gas and the dust differ slightly.
The difference in the motion between the gas and the dust is parameterized by
\begin{eqnarray}
v\sub{\phi}=v\sub{K}\left(1-\eta\right),
\end{eqnarray}
\begin{eqnarray}
\eta = -\frac{1}{2}\left(\frac{\cs}{v\sub{K}}\right)^{2}\frac{d\ln P}{d \ln r}
=-\frac{\Hgas^{2}}{2}\frac{d\ln P}{d \ln r},
\label{eq:eta0}
\end{eqnarray}
where $v\sub{\phi}$ and $v\sub{K}$ are the azimuthal and Kepler velocities, respectively, and $P$ is the gas pressure.
By using $d\ln P/d\ln r=d\ln\left(\Sigma\sub{gas}T/H\sub{gas}\right)/d\ln r$, and substituting to equations (\ref{eq:temp}), (\ref{eq:aspectratio}), and (\ref{eq:sigmag}), we obtain
\begin{eqnarray}
\frac{d\ln P}{d\ln r}=-\frac{24}{7}.
\end{eqnarray}
Therefore, the radial dependence of $\eta$ can be written as
\begin{eqnarray}
\eta\simeq 3.9\times10^{-3}\Mref^{-1}\tenau^{4/7}.
\label{eq:eta}
\end{eqnarray}

%%%%%%%%%%%%%%%%%%%%%%%%%%%%%%%%%%%%%%%%%%%%%%%

\subsection{Dust disk model}\label{dustdisk}

%%%%%%%%%%%%%%%%%%%%%%%%%%%%%%%%%%%%%%%%%%%%%%%

In our model, we assume analytical expressions for the dust disk that are derived by \citet{tsu17}, in which the Epstein law is assumed.
Here, we briefly summarize the derivation of the structures of the dust disk.

The evolution of the surface density of the dust  $\Sigma\sub{dust}$ and the mass of a dust particle $m\sub{dust}$ are described as follows:
\begin{eqnarray}
\frac{\partial\Sigma\sub{dust}}{\partial t}+\frac{1}{r}\frac{\partial}{\partial r}\left(r\vrdust\Sigma\sub{dust}\right)=0,\label{eq:dsdt1}\\
\frac{\partial m\sub{dust}}{\partial t}+\vrdust\frac{\partial m\sub{dust}}{\partial r}=\frac{m\sub{dust}}{t\sub{coll}},\label{eq:dmdt1}
\end{eqnarray}
where $\vrdust$ and $t\sub{coll}$ are the radial velocity of the dust particles and the collision time, respectively.
Because we assume the steady state, the time-derivative terms can be ignored (i.e., $\partial/\partial t=0$), and therefore equations (\ref{eq:dsdt1}) and (\ref{eq:dmdt1}) become
\begin{eqnarray}
\frac{1}{r}\frac{\partial}{\partial r}\left(r\vrdust\Sigma\sub{dust}\right)=0,\label{eq:dsdt2}\\
\vrdust\frac{\partial m\sub{dust}}{\partial r}=\frac{m\sub{dust}}{t\sub{coll}}.\label{eq:dmdt2}
\end{eqnarray}
Upon integrating equation (\ref{eq:dsdt2}), we see that $r\left|\vrdust\right|\Sigma\sub{dust}$ is constant, and this term becomes the mass flux when it is multiplied by $2\pi$.
Therefore, we obtain the following equation:
\begin{eqnarray}
2\pi r\left|\vrdust\right|\Sigma\sub{dust}=\Mddot. \label{eq:dsdt3}
\end{eqnarray}

The mass of a single dust particle is
\begin{eqnarray}
m\sub{dust}=\frac{4\pi}{3}\adust^{3}\rhoint
\label{eq:mdust}
\end{eqnarray}
where $\adust$ and $\rhoint$ are the radius and the internal density of the dust particle, respectively.
The collision time in equation (\ref{eq:dmdt2}) is
\begin{eqnarray}
t\sub{coll}=\frac{1}{4\pi \adust^{2}n\sub{dust}\Delta v\sub{d}}
\label{eq:tcoll}
\end{eqnarray}
where $n\sub{dust}$ and $\Delta v\sub{d}$ are the dust number density and the relative velocity, respectively, and $n\sub{dust}$ can be described as
\begin{eqnarray}
n\sub{dust}=\frac{\Sigma\sub{dust}}{\sqrt{2\pi}H\sub{dust} m\sub{dust}}.
\label{eq:ndust}
\end{eqnarray}
In this equation, $H\sub{dust}$ is the scale height of the dust disk.
In addition we assume
\begin{eqnarray}
\vrdust=-2\eta v\sub{K}\St.
\label{eq:vrdust}
\end{eqnarray}
$\St$ is the Stokes number that is defined by
\begin{eqnarray}
\St \equiv t\sub{stop}\Omega,
\label{eq:stokes0}
\end{eqnarray}
where $t\sub{stop}$ is the stopping time due to the gas drag force.
The stopping time is expressed as follows:
\begin{eqnarray}
t\sub{stop}=
\left\{
\begin{array}{ll}
\displaystyle{\frac{\rhoint\adust}{\rho\sub{gas}v\sub{th}}} & \left(\adust<\frac{9}{4}\mfp,\,{\rm Epstein\,drag\,law}\right)\\
\displaystyle{\frac{4\rhoint\adust^{2}}{9\rho\sub{gas}v\sub{th}\mfp}} & \left(\adust>\frac{9}{4}\mfp,\,{\rm Stokes\,drag\,law}\right)
\end{array}
\right.
\label{eq:tstop}
\end{eqnarray}
where $\rho\sub{gas}$, $v\sub{th}$, and $\mfp$ are the gas density, thermal velocity, and the mean free path, respectively.

By using equations (\ref{eq:eta0}), (\ref{eq:mdust}), (\ref{eq:tcoll}), (\ref{eq:ndust}), and (\ref{eq:vrdust}), equation (\ref{eq:dmdt2}) can be rewritten as the equation for the dust size as follows:
\begin{eqnarray}
\frac{\partial\adust}{\partial r}=-\left(\frac{1}{2\pi}\right)^{3/2}\frac{v\sub{K}^{2}\Mddot}{r\cs^{4}\left(d\ln P/d\ln r\right)^{2}\rhoint}\left(\frac{\Delta v}{H\sub{dust}\St^{2}}\right).
\end{eqnarray}
In the Epstein drag regime, the relative velocity and the dust scale height are described as
\begin{eqnarray}
\Delta v\sub{d}&=&\sqrt{\alpha\cs^{2}\St},\label{eq:deltav}\\
H\sub{dust}&\simeq&\left(1+\frac{\St}{\alpha}\right)^{-1/2}H\sub{gas}\simeq\left(\frac{\St}{\alpha}\right)^{-1/2}H\sub{gas}\label{eq:hdhg}
\end{eqnarray}
\citep[e.g.,][]{oc07,yl07}, and therefore the radial size distribution of the dust particle can be expressed as
\begin{eqnarray}
\frac{\partial\adust}{\partial r}=-\left(\frac{m\sub{g}^{2} v\sub{K}^{3}\Mddot\Sigma\sub{gas}}{\left(2\pi^{5}\right)^{1/2}\kB^{2}T^{2}\left(d\ln P/d\ln r\right)^{2}\rhoint^{2}}\right)r^{-2}\adust^{-1}.
\label{eq:dadr1}
\end{eqnarray}

As described by equations (\ref{eq:temp}) and (\ref{eq:sigmag}), the profiles of the temperature and the gas surface density are described by the power law as $\Sigma\sub{gas}=\Sigma\sub{0}r^{-12/7}$ and $T= T_{0}r^{-3/7}$.
The Kepler velocity $v\sub{K}=r\Omega\sub{K}$ is also written as $v\sub{K}= v\sub{K,0}r^{-1/2}$.
Therefore equation (\ref{eq:dadr1}) becomes
\begin{eqnarray}
\frac{\partial\adust}{\partial t}=A r^{-61/14}\adust^{-1},
\label{eq:dadr2}
\end{eqnarray}
where A is a negative constant defined as
\begin{eqnarray}
A\equiv -\frac{m\sub{g}^{2} v\sub{K,0}^{3}\Mddot\Sigma\sub{0}}{\left(2\pi^{5}\right)^{1/2}\kB^{2}T_{0}^{2}\left(d\ln P/d\ln r\right)^{2}\rhoint^{2}}.
\label{eq:A}
\end{eqnarray}
By integrating equation (\ref{eq:dadr2}), we obtain
\begin{eqnarray}
\adust=\left(-\frac{28}{47}Ar^{-47/14}+C\right)^{1/2},
\end{eqnarray}
where $C$ is an integral constant.
Because the index of the power law is negative, the dust size asymptotically approaches in $r\rightarrow 0$,
\begin{eqnarray}
\adust=\left(-\frac{28}{47}A\right)^{1/2}r^{-47/28}.
\label{eq:adust0}
\end{eqnarray}
Equations (\ref{eq:A}) and (\ref{eq:adust0}) tell us that the dependence of the dust size is $\adust\propto r^{-47/28}\Mgdot^{1/2}\Mstar\rhoint^{-1}$ when the dust accretion rate is proportional to the gas accretion rate, which we assume in equation (\ref{eq:MgdotMddot}) in Section \ref{results}.
Normalizing equation (\ref{eq:adust0}) with the values at 10 au, we obtain the analytic solution for the dust size in our disk model as
\begin{eqnarray}
\adust\simeq3.8\times10^{2}\tenau^{-47/28}\Mgdotref^{1/2}\nonumber\\
\Mref\rhoref^{-1}\,{\rm cm}.
\label{eq:adust}
\end{eqnarray}

As we describe in a later section (equation (\ref{eq:Stokes})), the Stokes number in the Epstein drag regime is written as $\St=\pi\rhoint\adust/2\Sigma\sub{gas}$.
Because the dependences of the dust size and the gas surface density are $\Sigma\sub{gas}\propto r^{-12/7}\Mstar^{1/2}$ and $\adust\propto r^{-47/28}\Mgdot^{1/2}\Mstar\rhoint^{-1}$, the dependence of the Stokes number is $\St\propto r^{1/28}\Mgdot^{1/2}\Mstar^{1/2}$.
From equations (\ref{eq:cs}), (\ref{eq:alpha}), and (\ref{eq:deltav}), we obtain the result that the dependence of the relative velocity is $\Delta v\sub{d}\propto r^{1/8}\Mgdot^{3/4}\Mstar^{1/4}$.
In addition, combining equations (\ref{eq:eta}) and (\ref{eq:vrdust}) yields $\vrdust\propto r^{3/28}\Mgdot^{1/2}$.
We can then obtain the dependence of the dust surface density from equation (\ref{eq:dsdt3}) as $\Sigma\sub{dust}\propto r^{-31/28}\Mgdot^{1/2}$.
Calculating the constant factors and normalizing as in equation (\ref{eq:adust}), we finally obtain the analytic solution for the surface density of the dust component in the steady state as follows:
\begin{eqnarray}
\Sigma\sub{dust}\simeq1.0\tenau^{-31/28}\Mgdotref^{1/2}\,\gcmtw.
\label{eq:sigmad}
\end{eqnarray}
This analytic solutions are derived assuming equation (\ref{eq:vrdust}), meaning that the dust has grown and radial drift already has begun in the disk.
\citet{tsu17} showed with 1D simulations that the growth timescale is short enough in $r\lesssim100$~au and the dust profiles quickly settles into this steady state solution within $t<10^{5}$ years.
Thus, this solution well describes the dust structure of the disk for $r\lesssim 100$~au where we investigate in this paper.

In addition, we assumed that the dust and gas are supplied at the edge of the disk.
This assumption is not standard for 1D simulations of viscous accretion disks with dust evolution \citep[e.g.,][]{hg05,tm18} in which the mass is added at the centrifugal radius.
However, recent 3D magnetohydrodynamical simulations show the formation of the outflow in the upper envelope and formation of the flattened envelope structure so-called pseudo-disk surrounding disk\citep{mac11,tom15,tsu15,tsu15b,mas16,wur16}.
These structures may suppress the mass accretion from the upper envelope and enhance mass accretion from the disk edge.
Thus, this assumption is valid for the disk in class 0/I YSOs.

%%%%%%%%%%%%%%%%%%%%%%%%%%%%%%%%%%%%%%%%%%%%%%%
%%%%%%%%%%%%%%%%%%%%%%%%%%%%%%%%%%%%%%%%%%%%%%%

\subsection{Stokes number}\label{pebbleaccretion}

%%%%%%%%%%%%%%%%%%%%%%%%%%%%%%%%%%%%%%%%%%%%%%%
%%%%%%%%%%%%%%%%%%%%%%%%%%%%%%%%%%%%%%%%%%%%%%%

From this section we derive the timescale of the pebble accretion in our disk model.
To achieve this, first we calculate several important parameters such as the stopping time and Stokes number, after which we estimate the modes of the pebble accretion and their timescales.

The Stokes number describes how the dust particles and pebbles are coupled to the gas flow.
When the size of the dust particle is small and the mean free path is large due to the low gas density (e.g., in the outer region of the disk), the stopping time is determined by the Epstein drag law.
In contrast, when the dust grows larger and the mean free path is small due to the high gas density, the stopping time is determined by the Stokes drag law.
By using equations (\ref{eq:stokes0}) and (\ref{eq:tstop}), the Stokes number is rewritten as
\begin{eqnarray}
\St=
\left\{
\begin{array}{ll}
\displaystyle{\frac{\pi\rhoint\adust}{2\Sigma\sub{gas}}} & \left({\rm Epstein}\right)\\
\displaystyle{\frac{2\pi\rhoint\adust^{2}}{9\Sigma\sub{gas}\mfp}} & \left({\rm Stokes}\right).
\end{array}
\right.
\label{eq:Stokes}
\end{eqnarray}

The thermal velocity is expressed as $v\sub{th}=\sqrt{8/\pi}\cs$, and the scale height of the gas disk is $H\sub{gas}=\cs/\Omega$.
The relation between the gas density at the midplane of the disk and the surface density of the gas is
\begin{eqnarray}
\rho\sub{gas}=\frac{\Sigma\sub{gas}}{\sqrt{2\pi}H\sub{gas}}.
\label{eq:rhogas}
\end{eqnarray}

The mean free path can be described as follows;
\begin{eqnarray}
\mfp=\frac{m\sub{gas}}{\sigma\sub{mol}\rho\sub{gas}}&=&\frac{m\sub{gas}}{\sigma\sub{mol}}\frac{\sqrt{2\pi}H\sub{gas}}{\Sigma\sub{gas}} \nonumber\\
&\simeq&5.2\times10^{1}\Mref^{-1}\tenau^{3}\,{\rm cm},
\label{eq:mfp}
\end{eqnarray}
where $\sigma\sub{mol}$ ($\simeq 2\times10^{-15}\,{\rm cm^{2}}$) is the collisional cross section of the gas molecules.

In the Epstein regime, using equation (\ref{eq:sigmag}), the Stokes number is given by
\begin{eqnarray}
\St \simeq 2.3\times10^{-4}\rhoref \Mref^{-1/2}\nonumber\\
\left(\frac{\adust}{1.0\,{\rm cm}}\right)\tenau^{12/7}.
\label{eq:StokesEpstein}
\end{eqnarray}
By substituting equation (\ref{eq:adust}) into this equation, we obtain the following expression;
\begin{eqnarray}
\St \simeq 8.9\times10^{-2}\Mref^{1/2} \Mgdotref^{1/2} \tenau^{1/28}.
\label{eq:StokesEpstein2}
\end{eqnarray}
The radial dependence of the Stokes number in the Epstein regime is very weak, $\St\propto r^{1/28}$.

In the Stokes regime, using equation (\ref{eq:sigmag}), the Stokes number is given by
\begin{eqnarray}
\St = 2.0\times10^{-6} \rhoref \nonumber\\
\left(\frac{\adust}{1.0\,{\rm cm}}\right)^{2} \Mref^{1/2} \tenau^{-9/7}.
\label{eq:StokesStokes}
\end{eqnarray}
The analytical expression of $\adust$ (equation (\ref{eq:adust})) is derived by assuming the Epstein regime, and therefore it cannot be used for the Stokes regime.
Therefore we left $\adust$ in equation (\ref{eq:StokesStokes}).
Equation (\ref{eq:adust}) indicates that the radius of the dust particles becomes approximately 400~cm around 10~au where the drag law changes from the Epstein to Stokes regime.

From equations (\ref{eq:adust}) and (\ref{eq:mfp}), the transition radius from the Epstein to the Stokes regime $r\sub{ES}$ can be calculated as
\begin{eqnarray}
r\sub{ES}\simeq 13 \Mref^{56/131} \Mgdotref^{14/131} \nonumber\\
\rhoref^{-28/131}\,{\rm au}.
\label{eq:rES}
\end{eqnarray}
Our primary focus is $r>10$ au and we only consider the Epstein regime in this paper.

%%%%%%%%%%%%%%%%%%%%%%%%%%%%%%%%%%%%%%%%%%%%%%%
%%%%%%%%%%%%%%%%%%%%%%%%%%%%%%%%%%%%%%%%%%%%%%%

\subsection{Accretion timescale}\label{accretion}

%%%%%%%%%%%%%%%%%%%%%%%%%%%%%%%%%%%%%%%%%%%%%%%
%%%%%%%%%%%%%%%%%%%%%%%%%%%%%%%%%%%%%%%%%%%%%%%

Next we estimate the timescale of the pebble accretion.
The accretion mode of the pebbles are divided into two regimes, the 3D regime and the 2D regime \citep[e.g.,][]{ida16}.
In addition, the pebble accretion mode is also divided into Bondi and Hill regimes, mainly depending on the mass of planetesimals/planetary embryos \citep{gui14}.
In this section, we denote the accretion timescales in these regimes and evaluate which regimes will be dominant in our disk model.

%%%%%%%%%%%%%%%%%%%%%%%%%%%%%%%%%%%%%%%%%%%%%%%

\subsubsection{Bondi/Hill accretion modes}\label{BH}

%%%%%%%%%%%%%%%%%%%%%%%%%%%%%%%%%%%%%%%%%%%%%%%

When the mass of the planetary embryo is small, accretion occurs in the Bondi regime, and then it changes to the Hill regime as the mass increases \citep{gui14}.
In the Bondi regime, the equivalent collisional cross section of a planetesimal/planetary embryo is written as
\begin{eqnarray}
b\simeq\kappa\sqrt{\frac{12\St R\sub{H}^{3}}{\eta' r}},
\label{eq:bbondi}
\end{eqnarray}
and in the Hill regime it is written as
\begin{eqnarray}
b\simeq2\kappa\St^{1/3}R\sub{H},
\label{eq:bhill}
\end{eqnarray}
where $\eta'=\chi\eta$ and $\chi=\sqrt{1+4\St^{2}}/\left(1+\St^{2}\right)$ \citep{ida16}.
$R\sub{H}$ is the Hill radius, which is given by
\begin{eqnarray}
R\sub{H}=r\left(\frac{M}{3\Mstar}\right)^{1/3},
\end{eqnarray}
where $M$ is the mass of a planetary embryo.
In these equations, $\kappa$ is the reduction factor, proposed by \citet{ok12}:
\begin{eqnarray}
\kappa=\exp\left(-\left(\frac{\St}{{\rm min}\left(2,\,\St^{\ast}\right)}\right)^{0.65}\right)
\label{eq:kappa}
\end{eqnarray}
\begin{eqnarray}
\St^{\ast}=4\left(\frac{M}{\Mstar}\right)\frac{1}{\eta^{3}}.
\label{eq:stast}
\end{eqnarray}
In the case of relatively small $\St$ and large $M$, $\kappa$ can be approximated to unity, thus this factor can be omitted \citep{ida16}.
However, this factor cannot be neglected when $\St$ is relatively large and/or the mass of the planetesimal/planetary embryo is small.

The transition from the Bondi to the Hill regime occurs when the following condition is satisfied:
\begin{eqnarray}
\sqrt{\frac{3\St^{1/3}R\sub{H}}{\eta'r}}\sim1,
\end{eqnarray}
In this equation, $\chi\sim1$ and $\eta'\sim\eta$ for $\St <1$.
From these criteria, we obtain the transition mass between the Bondi and Hill regimes as follows:
\begin{eqnarray}
M\sub{BH}&\simeq& \frac{\eta^{3}}{9\,\St}\Mstar \nonumber\\
&\simeq&2.2\times10^{-2}\left(\frac{\St}{0.1}\right)^{-1}\nonumber\\
&&\Mref^{-3}\tenau^{12/7}\,\MEarth.
\end{eqnarray}

By substituting equation (\ref{eq:StokesEpstein2}), this can be rewritten as
\begin{eqnarray}
M\sub{BH}\simeq2.5\times10^{-2}\Mref^{-7/2}\Mgdotref^{-1/2}\nonumber\\
\tenau^{47/28}\,\MEarth.
\label{eq:MBH}
\end{eqnarray}
This shows that the transition from the Bondi regime to the Hill regime occurs when the mass of the planetary embryo becomes $\sim \! 0.025\,\MEarth$ at 10 au in the Epstein regime.

From equation (\ref{eq:bbondi}), the equivalent accretion cross section in the Bondi regime can be written as
\begin{eqnarray}
b\sub{Bondi}\simeq2\kappa\left(\frac{\St}{\eta}\frac{M}{\Mstar}\right)^{1/2}r.
\end{eqnarray}
When we assume $\St\sim0.1$ and $\eta\sim10^{-3}$, $b\sub{Bondi}$ is
\begin{eqnarray}
b\sub{Bondi}\simeq5.1\times10^{5}\kappa\left(\frac{\St}{0.1}\right)^{1/2}\left(\frac{\eta}{10^{-3}}\right)^{-1/2}\left(\frac{M}{10^{-4}\MEarth}\right)^{1/2}\nonumber\\
\Mref^{-1/2}\tenau\,{\rm km}.
\end{eqnarray}
In the Hill regime, the equivalent accretion cross section is written as
\begin{eqnarray}
b\sub{Hill}\simeq2\kappa\St^{1/3}R\sub{H}
\end{eqnarray}
by equation (\ref{eq:bhill}) and this can be expressed as
\begin{eqnarray}
b\sub{Hill}\simeq9.3\times10^{5}\kappa\left(\frac{\St}{0.1}\right)^{1/3}\left(\frac{M}{10^{-4}\MEarth}\right)^{1/3}\Mref^{-1/3}\nonumber\\
\tenau\,{\rm km}.
\end{eqnarray}
These expressions show that the equivalent accretion cross sections in both the Bondi and Hill regimes are much larger than the physical radius of planetesimals or planetary embryos.
Therefore, the equivalent accretion cross sections in the Bondi and Hill regimes are always larger than the physical cross section.

%%%%%%%%%%%%%%%%%%%%%%%%%%%%%%%%%%%%%%%%%%%%%%%

\subsubsection{3D accretion mode}\label{3Daccretion}

%%%%%%%%%%%%%%%%%%%%%%%%%%%%%%%%%%%%%%%%%%%%%%%

When the scale height of the dust/pebble disk is larger compared to the collision cross section of a planetesimal or a planetary embryo, and the planetary embryos are not so massive, pebble accretion occurs in the 3D regime.
Therefore, the accretion mode in the outer region of the disk tend to be the 3D regime.
In the case of isotropic pebble flux, the accretion rate can be written as
\begin{eqnarray}
\dot{M}=\pi b^{2}\rho\sub{dust}\Delta v
\end{eqnarray}
where $\rho\sub{dust}$ and $\Delta v$ are the spatial density of the pebbles, and the relative velocity between the pebbles and a planetesimal or a planetary embryo.
Considering the relation between the spatial density and the surface density of the pebbles, $\rho\sub{dust}=\Sigma\sub{dust}/\sqrt{2\pi}H\sub{dust}$, which is equivalent to equation (\ref{eq:ndust}), we obtain
\begin{eqnarray}
\dot{M}=\pi b^{2}\frac{\Sigma\sub{dust}}{\sqrt{2\pi}H\sub{dust}}\Delta v.
\label{eq:Mdot1}
\end{eqnarray}

The accretion timescale is defined as $t\sub{acc}\equiv M/\dot{M}$.
The relative velocity $\Delta v$ is given by
\begin{eqnarray}
\Delta v\simeq\left(1+\frac{3b}{2\eta r}\right)\eta v\sub{K}.
\end{eqnarray}
In the Bondi regime, the accretion cross section $b$ is given by equation (\ref{eq:bbondi}).
Therefore, the accretion timescale can be expressed as
\begin{eqnarray}
\taccth\simeq\sqrt{\frac{2}{\pi}}\frac{\Mstar \Hgas\alpha^{1/2}}{4r^{2}\Sigma\sub{dust}\Omega\sub{K}\St^{3/2}\kappa^{2}}.
\label{eq:taccthbondi}
\end{eqnarray}
In the Hill regime, $b$ is given by equation (\ref{eq:bhill}), thus $\taccth$ is
\begin{eqnarray}
\taccth\simeq\sqrt{\frac{2}{\pi}}\frac{\Mstar \Hgas\alpha^{1/2}}{4r^{2}\Sigma\sub{dust}\Omega\sub{K}\St^{3/2}\kappa^{3}}.
\label{eq:taccthhill}
\end{eqnarray}
As pointed out in \citet{ida16}, the accretion timescale in the 3D regime is identical in both the Bondi and Hill regimes except for the reduction factor $\kappa$, which becomes important when $\St$ is large and the mass of the embryo is small.
As described in section \ref{BH}, the accretion occurs in the Bondi regime during the mass of the embryo is small (equation (\ref{eq:MBH})).
Thus the early phase of the growth of the embryo is in the Bondi regime that is described by equation (\ref{eq:taccthbondi}).

By substituting equations (\ref{eq:aspectratio}), (\ref{eq:Omega}), (\ref{eq:alpha}), (\ref{eq:sigmad}), and assuming that the drag law is in the Epstein regime (equation (\ref{eq:StokesEpstein2})), equation (\ref{eq:taccthbondi}) is rewritten as
\begin{eqnarray}
\taccth&\simeq&
9.4\times10^{5}\kappa^{-2}\Mref^{-3/4}\Mddotref^{-1}\nonumber\\
&&\Mgdotref^{1/4}\tenau^{65/56}\,{\rm yr}.
\label{eq:taccthEpstein}
\end{eqnarray}
This is the accretion timescale in the 3D and Epstein regime (and in the Bondi regime).
In the Hill regime, $\kappa^{-2}$ in equation (\ref{eq:taccthEpstein}) is replaced by $\kappa^{-3}$.

In the Stokes regime, the Stokes number is expressed by equation (\ref{eq:StokesStokes}), and therefore the accretion timescale is
\begin{eqnarray}
\taccth\simeq5.0\times10^{5}\kappa^{-2}\Mref^{-3/4}\Mgdotref^{1/2}\nonumber\\
\Mddotref^{-1}\rhoref^{-1/2}\nonumber\\
\left(\frac{\adust}{400\,{\rm cm}}\right)^{-1}\tenau^{51/28}\,{\rm yr}.
\label{eq:tacc3dStokes}
\end{eqnarray}
This is the accretion timescale in the 3D and Stokes regime.

%%%%%%%%%%%%%%%%%%%%%%%%%%%%%%%%%%%%%%%%%%%%%%%

\subsubsection{2D accretion mode}\label{2Daccretion}

%%%%%%%%%%%%%%%%%%%%%%%%%%%%%%%%%%%%%%%%%%%%%%%

In the inner region of the disk, the dust scale height becomes relatively thin.
In addition, the cross section of the accretion becomes larger as the planetary embryos grow, and thus the accretion mode changes from the 3D mode to the 2D mode.

In the 2D mode, the accretion rate is expressed by the following equation \citep{ida16}:
\begin{eqnarray}
\dot{M}=2b\Sigma\sub{dust}\Delta v.
\end{eqnarray}
Calculating the same manner as in Section \ref{3Daccretion}, we obtain 
\begin{eqnarray}
\tacctw\simeq8.8\times10^{4}\kappa^{-2}\Mref^{-1/3}\Mddotref^{-1}\nonumber\\
\left(\frac{\St}{0.1}\right)^{1/3}\left(\frac{M}{0.1\MEarth}\right)^{1/3}\tenau^{4/7}\,{\rm yr},
\end{eqnarray}
where $\MEarth$ is Earth's mass.
Here we assume that the collisional cross section $b$ is determined by the Bondi regime (see Section \ref{2D3D}).
When the drag law is the Epstein regime, we can substitute equation (\ref{eq:StokesEpstein2}) and obtain
\begin{eqnarray}
\tacctw\simeq8.5\times10^{4}\kappa^{-2}\Mref^{-1/6}\Mgdotref^{1/6}\nonumber\\
\Mddotref^{-1}\left(\frac{M}{0.1\MEarth}\right)^{1/3}\tenau^{7/12}\,{\rm yr}.
\end{eqnarray}
This is the accretion timescale in the 2D and Epstein regime.
In the Stokes regime, the accretion timescale is denoted as
\begin{eqnarray}
\tacctw\simeq1.3\times10^{5}\kappa^{-2}\Mref^{-1/6}\Mddotref^{-1}\nonumber\\
\rhoref^{1/3}\left(\frac{\adust}{400\,{\rm cm}}\right)^{2/3}\nonumber\\
\left(\frac{M}{0.1\,\MEarth}\right)^{1/3}\tenau^{1/7}\,{\rm yr}.
\end{eqnarray}
This is the accretion timescale in the 2D and Stokes regime.

%%%%%%%%%%%%%%%%%%%%%%%%%%%%%%%%%%%%%%%%%%%%%%%

\subsubsection{Transition of 2D/3D accretion modes}\label{2D3D}

%%%%%%%%%%%%%%%%%%%%%%%%%%%%%%%%%%%%%%%%%%%%%%%

The transition from the 3D accretion mode to the 2D accretion mode occurs when
\begin{eqnarray}
\hat{b}\sim\sqrt{\frac{8}{\pi}}\Hdust
\label{eq:2d3dtransition}
\end{eqnarray}
is satisfied \citep{ida16}, where $\hat{b}=b/r$ is a normalized cross section.
From equations (\ref{eq:hdhg}) and (\ref{eq:2d3dtransition}), the transition mass of the planetary embryo can be written as
\begin{eqnarray}
M\sub{2D3D}\simeq\frac{3\left(\sqrt{2/\pi}\Hdust\right)^{3}}{\St}.
\end{eqnarray}
From equations (\ref{eq:alpha}) and (\ref{eq:StokesEpstein2}), the transition mass is obtained as
\begin{eqnarray}
M\sub{2D3D}\simeq154\Mref^{-7/4}\Mgdotref^{1/4}\nonumber\\
\tenau^{97/56}\,\MEarth.
\end{eqnarray}
In our disk model, the transition mass of the planetary embryo is very large, and therefore the pebble accretion mode is dominated by the 3D accretion mode.
This transition mass is much larger than the transition mass between Bondi and Hill regime (see, equation (\ref{eq:MBH})).
Therefore the 2D accretion mode always occurs in the Hill regime \citep{ida16}.

In addition, as shown in equation (\ref{eq:rES}), the transition from the Epstein to the Stokes drag law occurs at $\sim\!10\,{\rm au}$.
Thus, the pebble accretion mode in our fiducial model is the 3D and Epstein regime, and its timescale is expressed by equation (\ref{eq:taccthEpstein}).

%%%%%%%%%%%%%%%%%%%%%%%%%%%%%%%%%%%%%%%%%%%%%%%
%%%%%%%%%%%%%%%%%%%%%%%%%%%%%%%%%%%%%%%%%%%%%%%
%%%%%%%%%%%%%%%%%%%%%%%%%%%%%%%%%%%%%%%%%%%%%%%

\section{Results}\label{results}

%%%%%%%%%%%%%%%%%%%%%%%%%%%%%%%%%%%%%%%%%%%%%%%
%%%%%%%%%%%%%%%%%%%%%%%%%%%%%%%%%%%%%%%%%%%%%%%
%%%%%%%%%%%%%%%%%%%%%%%%%%%%%%%%%%%%%%%%%%%%%%%

%%%%%%%%%%%%%%%%%%%%%%%%%%%%%%%%%%%%%%%%%%%%%%%
%%%%%%%%%%%%%%%%%%%%%%%%%%%%%%%%%%%%%%%%%%%%%%%

\subsection{Accretion timescales}\label{results_tacc}

%%%%%%%%%%%%%%%%%%%%%%%%%%%%%%%%%%%%%%%%%%%%%%%
%%%%%%%%%%%%%%%%%%%%%%%%%%%%%%%%%%%%%%%%%%%%%%%

In our disk model, the accretion rate is denoted by equation (\ref{eq:taccthEpstein}), which is the accretion timescale in the 3D and Epstein regime as described in Section \ref{2D3D}.
Here, we assume that the dust accretion rate is related to the gas accretion rate as follows:
\begin{eqnarray}
\Mddot=f\Mgdot,
\label{eq:MgdotMddot}
\end{eqnarray}
where $f$ is the dust-to-gas mass ratio.
If we assume $\Mgdot=10^{-6}\Msun\,{\rm yr^{-1}}$ and $f=0.01$ as a fiducial model, the accretion timescale of equation (\ref{eq:taccthEpstein}) is rewritten as
\begin{eqnarray}
\taccth\simeq2.8\times10^{4}\kappa^{-2}\Mref^{-3/4}\tenau^{65/56}\,{\rm yr}.
\end{eqnarray}
Therefore, the pebble accretion timescale at $\sim\!10$ au can be shorter than or comparable to the typical lifetime of the class 0/I object, $t\sub{life}\sim5\times10^{5}$ yr, depending on the value of the reduction factor $\kappa$.

%%%%%%%%%%%%%%%%%%%%%%%%%%%%%%%%%%%%%%%%%%%%%%%
%%%%%%%%%%%%%%%%%%%%%%%%%%%%%%%%%%%%%%%%%%%%%%%

\subsection{Parameter dependence of the accretion timescale}\label{dependence}

%%%%%%%%%%%%%%%%%%%%%%%%%%%%%%%%%%%%%%%%%%%%%%%
%%%%%%%%%%%%%%%%%%%%%%%%%%%%%%%%%%%%%%%%%%%%%%%

Here we show the parameter dependences of the accretion timescale in the 3D and Epstein regime.
As expressed by equations (\ref{eq:taccthEpstein}) and (\ref{eq:MgdotMddot}), the accretion timescale depends on the stellar mass, accretion rates of the gas and dust, radial distance, and dust-to-gas mass ratio.
If we fix the stellar mass and assume the relation expressed by equation (\ref{eq:MgdotMddot}), the important parameters are the gas accretion rate, dust-to-gas mass ratio, and radial distance.
However, the reduction factor $\kappa$ given by equation (\ref{eq:kappa}) depends on $\St$ and the mass of the planetesimal/planetary embryo, so the mass is also the important parameter for the accretion timescale.

%%%%%%%%%%%%%%%%%%%%%%%%%%%%%%%%%%%%%%%%%%%%%%%

\subsubsection{Dependence on the gas accretion rate}\label{dependence_gasaccretion}

%%%%%%%%%%%%%%%%%%%%%%%%%%%%%%%%%%%%%%%%%%%%%%%

First we investigate the radial dependence of the accretion timescales with different gas accretion rates.
Because the accretion mode in the outer region of the disk is the 3D and Epstein regime, we plot equations (\ref{eq:taccthEpstein}) and (\ref{eq:MgdotMddot}).
Here we assume that $\Mstar=\Msun$, $\Mddot=f\Mgdot$, and we fix the dust-to-gas mass ratio as $f=0.01$.
The accretion timescale also depends on the mass of the planetesimal $M$ because the reduction factor $\kappa$ is the function of $M$.
Here we assume that $M=10^{-3}\MEarth$.
\begin{figure}
   \centering
   \includegraphics[]{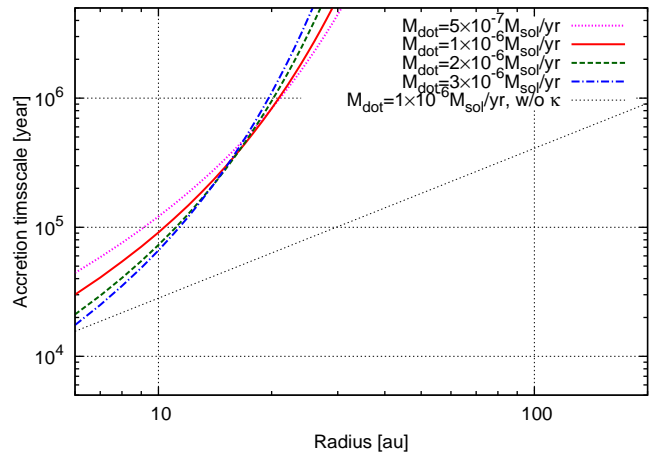} 
   \caption{
   Radial profiles of the accretion timescales with different gas accretion rates.
   The horizontal axis is the radial distance from the central star in units of astronomical unit (au), and the vertical axis is the accretion timescale in units of year.
   In this plot, we fixed $\Mstar=\Msun$, $\Mddot=f\Mgdot$, $f=0.01$, and $M=10^{-3}\MEarth$.
   The magenta dotted, red solid, green dashed, and blue dot-dashed lines correspond to the gas accretion rates $5\times10^{-7}\,\Msunyr$, $1\times10^{-6}\,\Msunyr$, $2\times10^{-6}\,\Msunyr$, and $3\times10^{-6}\,\Msunyr$, respectively.
   The black dotted line shows the accretion timescale when $\Mgdot=1\times10^{-6}\,\Msunyr$ and $\kappa=1$ are assumed.
   }
   \label{fig:taccretion}
\end{figure}

Fig. \ref{fig:taccretion} shows the radial dependence of the accretion timescale.
We also show the accretion timescale when $\kappa=1$ is assumed as a reference.
If the reduction factor is neglected, the radial dependence of the accretion timescale is $t\sub{acc}\propto r^{65/56}$, as described by equation (\ref{eq:taccthEpstein}), and therefore the timescale at $\sim\!100$ au is approximately ten times longer than that at $\sim\!10$ au.
The black dotted line in Fig. \ref{fig:taccretion} shows this relation.

However, the reduction is significant when $M$ is small and $\St$ is large.
From equations (\ref{eq:kappa}) and (\ref{eq:stast}), the reduction factor will be significant when $\St^{*}\lesssim2$.
The criterion mass can be written as $M\simeq(1/2)\eta^{3}\Mstar$, and this can be rewritten as
\begin{eqnarray}
M\simeq 2.5\times10^{-3}\Mref^{-2}\tenau^{12/7}\MEarth.
\label{eq:mcrit}
\end{eqnarray}
When the mass of the planetesimal/planetary embryo is smaller than this value, the effect of $\kappa$ is significant.
This effect is shown in Fig. \ref{fig:taccretion}; the accretion timescale described by the red solid line becomes larger than the black dotted line that $\kappa$ is ignored.
Equation (\ref{eq:mcrit}) also shows that the critical mass increases as the radial distance.
This is mainly caused by the radial dependence of $\eta$.
Therefore, the accretion timescale increases rapidly in the distant region.

The dependence of the accretion timescale on the gas accretion rate is not simple.
As shown in Fig. \ref{fig:taccretion}, the accretion timescale decreases with the increase of the gas accretion rate in the region within $\sim$ 20 au.
The dependence of the accretion timescale on the dust accretion rate is $\propto \Mddot^{-1}$ (equation (\ref{eq:taccthEpstein})), and the dependence of the timescale on the gas accretion rate is $\propto \Mgdot^{1/4}$.
Hence the resultant dependence of the timescale on the accretion rate becomes $\propto \Mgdot^{-3/4}$, when $\kappa$ is neglected.
In the distant region where $r\gtrsim$ 20 au, however, the accretion timescale slightly increases with the gas accretion rate.
This is caused by the dependence of $\St$ on the gas accretion rate.
The larger gas accretion rate causes the larger $\St$, and hence smaller reduction factor.
Since the critical mass is large in the distant region as shown by equation (\ref{eq:mcrit}), the effect of the reduction factor is significant.

%%%%%%%%%%%%%%%%%%%%%%%%%%%%%%%%%%%%%%%%%%%%%%%

\subsubsection{Dependence on the dust-to-gas mass ratio}\label{dependence_f}

%%%%%%%%%%%%%%%%%%%%%%%%%%%%%%%%%%%%%%%%%%%%%%%

In Section \ref{dependence_gasaccretion}, we assume that the dust-to-gas mass ratio in the disk is 0.01.
However, the dust-to-gas mass ratio varies from this in the different star forming region \citep{lis15}.
As shown in equation (\ref{eq:taccthEpstein}), the accretion timescale is inversely proportional to the dust accretion rate, and therefore the timescale becomes short as the dust-to-gas mass ratio increases.
\begin{figure}
   \centering
   \includegraphics[]{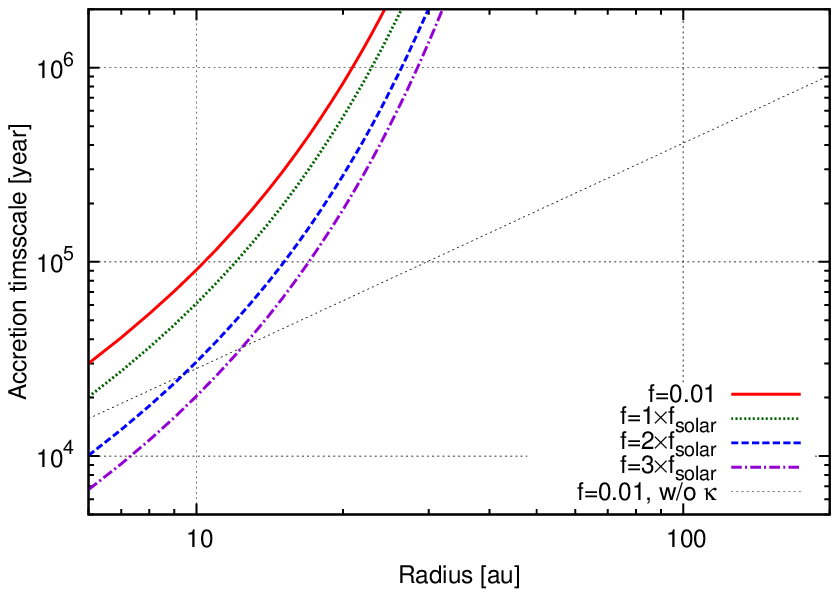} 
   \caption{
   Radial dependence of the accretion timescales with different values of the dust-to-gas mass ratio, $f$.
   The axes are the same as in Figure \ref{fig:taccretion}.
   In this plot, we fixed $\Mstar=\Msun$ and $\Mgdot=10^{-6}\,\Msunyr$, and assumed $\Mddot=f\Mgdot$.
   The red solid, green dotted, blue dashed, and violet dot-dashed lines correspond to the dust-to-gas mass ratios, $f=0.01$, $1\fsolar$, $2\fsolar$, and $3\fsolar$, respectively, and $\fsolar=0.0149$ is the dust-to-gas mass ratio in the solar nebula.
   The black dotted line shows the accretion timescale when $f=0.01$ and $\kappa=1$ are assumed.
   }
   \label{fig:taccretion2}
\end{figure}

Fig. \ref{fig:taccretion2} shows the radial dependence of the accretion timescale with different values of the dust-to-gas mass ratio.
We assume that $M=10^{-3}\MEarth$.
Here $\fsolar=0.0149$ is the dust-to-gas mass ratio for the solar system suggested by \citet{lod03}.
Because we fixed $\Mgdot=10^{-6}\,\Msunyr$, the accretion timescale is proportional to $t\sub{acc}\propto f^{-1}$.
Therefore, when the dust-to-gas mass ratio is doubled, the accretion timescale falls to one-half.
The general trend is similar to Fig. \ref{fig:taccretion}: the accretion timescale increases as the radial distance increases, because of the large value of $\eta$.
In this plot the gas accretion rate is fixed, and $\St$ does not depend on the dust-to-gas mass ratio.
Therefore the reduction factor does not depend on the value of the dust-to-gas mass ratio.

%%%%%%%%%%%%%%%%%%%%%%%%%%%%%%%%%%%%%%%%%%%%%%%

\subsubsection{Dependence on the embryo mass}\label{dependence_mass}

%%%%%%%%%%%%%%%%%%%%%%%%%%%%%%%%%%%%%%%%%%%%%%%

The accretion timescale does not depend on the embryo mass except for the reduction factor $\kappa$, and $\kappa$ gets close to unity when the mass of the embryo is large.
Therefore, in the case that the embryo mass is relatively large and $\St$ is small, $\kappa$ will be neglected.
\begin{figure}
   \centering
   \includegraphics[]{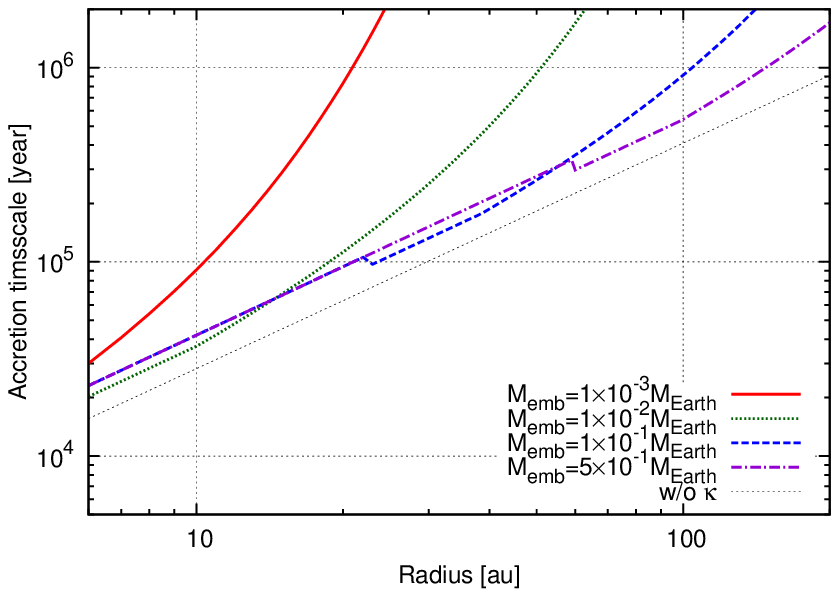} 
   \caption{
   Radial profiles of the accretion timescales with different masses of the embryo.
   The axes are the same as in Figure \ref{fig:taccretion}.
   In this plot, we fixed $\Mstar=\Msun$, $\Mgdot=1\times10^{-6}\,\Msunyr$, $\Mddot=f\Mgdot$, and $f=0.01$.
   The red solid, green dotted, blue dashed, and violet dot-dashed lines correspond to the embryo mass, $10^{-3}\MEarth$, $10^{-2}\MEarth$, $10^{-1}\MEarth$, and $5\times10^{-1}\MEarth$, respectively.
   The black dotted line shows the accretion timescale when $\kappa=1$ are assumed.
   }
   \label{fig:taccretion3}
\end{figure}
Fig. \ref{fig:taccretion3} shows the dependence of the accretion timescale on the embryo mass.
In case of the small embryo mass, the effect of the reduction factor is large and the accretion timescale largely deviates from the black dotted line that assumes $\kappa$ is unity.
When the embryo mass is large, the effect of the reduction factor becomes smaller, and the dependence of the accretion timescale on the radial distance becomes $\propto r^{65/56}$.
Fig. \ref{fig:taccretion3} clearly shows this dependence when the embryo mass is large.

In this plot, the accretion timescale does not completely match the value that assumes $\kappa=1$ even when the embryo mass is large, for example $M=5.0\times10^{-1}\MEarth$.
This is because the large value of $\St$ that is caused by the larger gas accretion rate.
In our model, the gas accretion rate is assumed to be much larger compared to the class II disks.
This leads to the $\St$ value about an order of magnitude larger the that of the class II disks \citep[e,g,][]{ida16}.
Therefore, the reduction factor does not completely reach unity, and certain amount of the effect of the reduction remains.

In Fig. \ref{fig:taccretion3}, small drops appear when $M=10^{-1}\MEarth$ and $M=5.0\times10^{-1}\MEarth$.
This is caused by the transition between Bondi and Hill regime.
From equation (\ref{eq:MBH}) we obtain the transition radial distance between the Bondi and Hill regime as follows:
\begin{eqnarray}
r\sub{BH}\simeq10\left(\frac{M}{2.5\times10^{-2}\MEarth}\right)^{28/47}\Mref^{98/47}\nonumber\\
\Mgdotref^{14/47}\,{\rm au}.
\end{eqnarray}
The transition occurs at $r\sub{BH}\sim5.8$~au for $M=10^{-2}\MEarth$, thus the accretion occurs in the Bondi regime in wide region of the disk.
However, $r\sub{BH}$ is $\sim23$ au and $\sim 60$~au when $M=10^{-1}\MEarth$ and $5.0\times10^{-1}\MEarth$, respectively.
In these cases, the region inner than $r\sub{BH}$ is the Hill regime, and the outer region is the Bondi regime.
As shown by equations (\ref{eq:taccthbondi}) and (\ref{eq:taccthhill}), the accretion timescales between the Bondi and Hill regime differ factor of $\kappa$.
This causes the small drops in the accretion timescales at $r\sub{BH}$.

%%%%%%%%%%%%%%%%%%%%%%%%%%%%%%%%%%%%%%%%%%%%%%%
%%%%%%%%%%%%%%%%%%%%%%%%%%%%%%%%%%%%%%%%%%%%%%%

\subsection{Growth of planets by the pebble accretion}\label{results_growth}

%%%%%%%%%%%%%%%%%%%%%%%%%%%%%%%%%%%%%%%%%%%%%%%
%%%%%%%%%%%%%%%%%%%%%%%%%%%%%%%%%%%%%%%%%%%%%%%

As mentioned in Section \ref{accretion} the definition of the accretion timescale is
\begin{eqnarray}
t\sub{acc}\equiv\frac{M}{\dot{M}}.
\label{eq:tacc}
\end{eqnarray}
If the accretion timescale does not depend on the embryo mass, i.e., $\kappa$ is assumed to be unity, can derive the mass growth of the planetary embryo as follows:
\begin{eqnarray}
M\!\left( t\right)/\Minit=\exp\!\left(t/t\sub{acc}\right),
\end{eqnarray}
where $\Minit$ is the initial mass of a planetesimal or a planetary embryo.
Using the lifetime of the class 0/I objects, $t\sub{life}$, the final mass of a grown-up protoplanet can be expressed by $M\!\left(t\sub{life}\right)$.
If $M\!\left(t\sub{life}\right)$ is sufficiently large, for example, $\sim\!10$ $\MEarth$, the gas accretes to the protoplanet rapidly, and the protoplanet evolves into a gas giant \citep[e.g.,][]{miz80,ste82,pol96,iko00}.
If the core of the gas giant is formed via the pebble accretion in the class 0/I disk within its lifetime, it is a promising pathway to the early formation of the gas giants in distant orbits.
By assuming $M\!\left(t\sub{life}\right)$ to form the gas giants, the required initial mass to form such a core is expressed by
\begin{eqnarray}
\Minit=10\,\MEarth \exp\!\left(-t\sub{life}/t\sub{acc}\right).
\label{eq:Minit}
\end{eqnarray}

As described in Section \ref{dependence}, however, $\kappa$ cannot be omitted when the mass of the embryo is small.
In addition, $\St$ is relatively large in our disk model due to the large gas accretion rate, the effect of $\kappa$ slightly remains even when the $M$ is large.
These effects cause that the accretion timescale depends on the embryo mass.
Therefore we integrate equation (\ref{eq:tacc}) for the disk lifetime to derive the required initial mass to form the core of the gas giants.
We set the final mass to 10~$\MEarth$, and integrate equation (\ref{eq:tacc}) inversely over the disk lifetime, $t\sub{life}$.

%%%%%%%%%%%%%%%%%%%%%%%%%%%%%%%%%%%%%%%%%%%%%%%

\subsubsection{Dependence on the gas accretion rate}\label{results_gasaccretion}

%%%%%%%%%%%%%%%%%%%%%%%%%%%%%%%%%%%%%%%%%%%%%%%

The required initial mass to form the core of the gas giant strongly depends on the accretion timescales and the required initial mass changes significantly.
First we show the dependence on the gas accretion rate.
\begin{figure}
   \centering
   \includegraphics[]{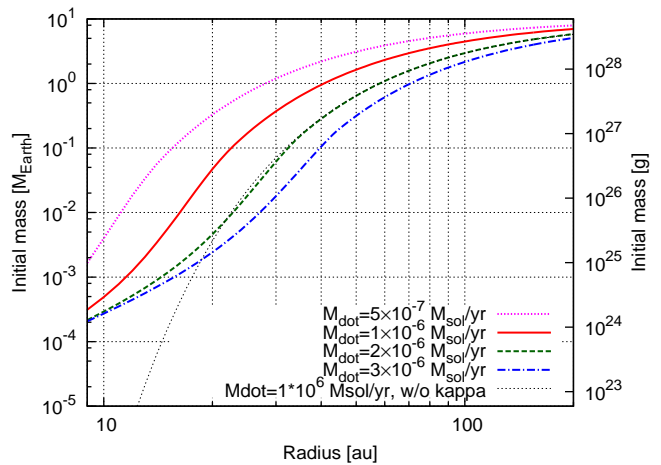} 
   \caption{
   Radial dependence of the required initial mass to form a 10 $\MEarth$ core within the disk lifetime with different gas accretion rates.
   The horizontal axis is the radial distance from the central star in units of au.
   The vertical axes are the required initial mass, the left-hand side axis is in units of $\MEarth$, and the right-hand side is in units of gram. 
   $\Mstar=\Msun$, $\Mddot=f\Mgdot$, $t\sub{life}=0.5$ Myr, and $f=0.01$ are assumed in this plot.
   The magenta dotted, red solid, green dashed, and blue dot-dashed lines correspond to the gas accretion rate, $5\times10^{-7}\,\Msunyr$, $1\times10^{-6}\,\Msunyr$, $2\times10^{-6}\,\Msunyr$, and $3\times10^{-6}\,\Msunyr$, respectively, and the blue dashed horizontal line shows $M=10^{-7}\,\MEarth$.
   The black dotted line shows the required initial mass when $1\times10^{-6}\,\Msunyr$ and $\kappa=1$ are assumed, and it corresponds to the plot of equation (\ref{eq:Minit}).
   }
   \label{fig:growth}
\end{figure}
Figure \ref{fig:growth} show the radial dependence of the required initial mass to form the 10 $\MEarth$ core, with the difference gas accretion rate assuming $t\sub{life}=$ 0.5 Myr, $\Mstar=\Msun$, $\Mddot=f\Mgdot$, and $f=0.01$, respectively.

When the gas accretion rate is $\Mgdot=10^{-6}\,\Msunyr$, the required initial mass is $\sim5.0\times10^{-4}\,\MEarth$ at 10~au.
Therefore, if the embryo whose mass is $\sim5.0\times10^{-4}\,\MEarth$ forms at 10~au, it can grow to the core of the gas giant within the lifetime of the class 0/I phases.
Since the accretion timescale is much longer in the distant region, the required initial mass have to be much larger.

The required initial mass depends on the gas accretion rate, and larger gas accretion rates result in a smaller initial mass.
However, since the reduction factor depends on the embryo mass exponentially, the required initial mass cannot be very small even in the case that the gas accretion rate is larger.
This effect can be seen in the inner region, $\sim$ 10--20~au: the required initial mass decreases as the increase of the gas accretion rate, but the amount of the decrease is not large compared to the outer region.

We also plot the required initial mass when $\kappa=1$ is assumed, as a reference.
When the effect of $\kappa$ is ignored, the accretion timescale does not depend on the embryo mass, so the required initial mass can be solved analytically, as denoted by equation (\ref{eq:Minit}).
This is plotted by black dotted line in Fig. \ref{fig:growth}.
When the required initial mass is large, for example in the distant region, the appearance of the lines are similar.
However they deviate significantly when the required initial mass is small.

According to the standard model of inside-out collapse of a gas cloud \citep{shu77}, the accretion rate is approximated by $\dot{M}\approx0.975\cs^{3}/G$, which is the so-called Shu accretion rate.
When we adopt 10 -- 20 K for the temperatures of typical star forming regions, the accretion rate becomes $\sim\!\left(1.6-4.5\right)\times 10^{-6}~\Msun\,{\rm yr^{-1}}$.
In addition, the Shu accretion rate is nearly the lower limit of the assumptions for the collapse of the gas cloud; it can be several tens of times larger than the Shu accretion rate, for example, $\dot{M}\approx 46.9\cs^{3}/G$ \citep[e.g.,][]{lar69,pen69,hun77}.
In this case a much larger accretion rate is possible, thus we investigate the value of required initial mass and its dependence when the gas accretion rate is large.
\begin{figure}
   \centering
   \includegraphics[]{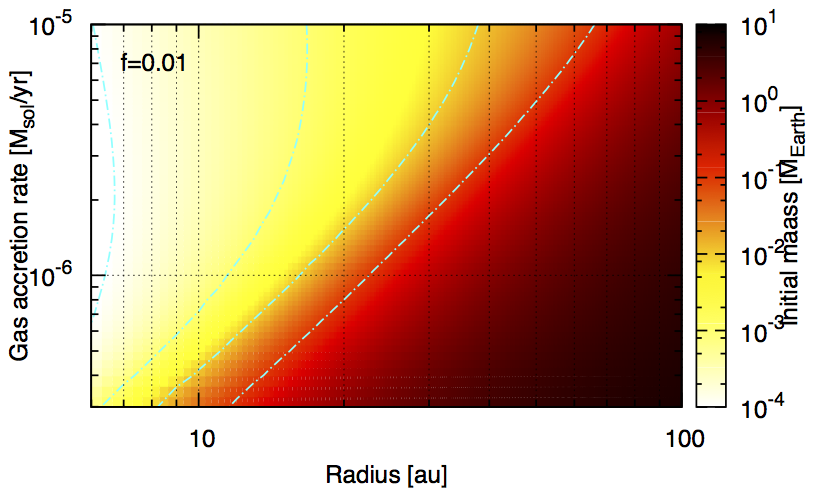} 
   \caption{
   Map of the required initial mass in a radius-gas accretion rate plane.
   The color bar corresponds to the required initial mass to form a 10 $\MEarth$ core.
   $\Mstar=\Msun$, $\Mddot=f\Mgdot$, $t\sub{life}=0.5$ Myr, and $f=0.01$ are assumed in this plot.
   The blue dashed lines shows the contour for a required initial mass.
   From left to right, $10^{-4}\MEarth$, $10^{-3}\MEarth$, $10^{-2}\MEarth$, and $10^{-1}\MEarth$, respectively.
   }
   \label{fig:initialmassmap}
\end{figure}

Figure \ref{fig:initialmassmap} shows the dependence of the initial mass on the radial distance and the gas accretion rate.
The blue dashed lines shows the contour for a required initial mass.
From left to right, $10^{-4}\MEarth$, $10^{-3}\MEarth$, $10^{-2}\MEarth$, and $10^{-1}\MEarth$, respectively.
It is shown that the larger gas accretion rate tends to reduce the resultant required initial mass.
However, the trend will break when the gas accretion rate is larger, especially for the smaller embryo mass.
This is caused by the dependence of the accretion timescale on the gas accretion rate that is described in Section \ref{dependence_gasaccretion}.

When the gas accretion rate is relatively small, the effect of the reduction factor is also small due to the small value of $\St$.
In this regime, the accretion timescale decreases as the gas accretion rate increases, and consequently the required initial mass decreases.

When the gas accretion rate becomes large, however, $\St$ also increases and the increase of the accretion timescale is significant.
In this regime, as the gas accretion rate increases, the increase of the accretion timescale due to the reduction factor will exceed the decrease of the accretion timescale due to the large gas accretion rate.
This effect appears in the contour of Fig. \ref{fig:initialmassmap}; the leftmost contour ($10^{-4}\MEarth$) is the curve convex to the right.
This result suggests that the larger gas accretion rate does not always help the efficient growth of the planetary embryo by the pebble accretion in the class 0/I phases.

%%%%%%%%%%%%%%%%%%%%%%%%%%%%%%%%%%%%%%%%%%%%%%%

\subsubsection{Dependence on the dust-to-gas mass ratio}\label{results_f}

%%%%%%%%%%%%%%%%%%%%%%%%%%%%%%%%%%%%%%%%%%%%%%%

As described in Section \ref{dependence_f}, it is expected that the required initial mass decreases when the dust-to-gas mass ratio is large.
\begin{figure}
   \centering
   \includegraphics[]{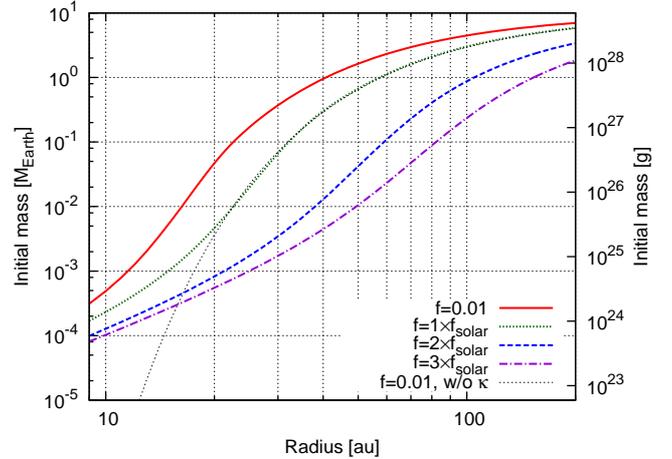} 
   \caption{
   Radial dependence of the required initial mass to form a 10 $\MEarth$ core within the disk lifetime with different values of the dust-to-gas mass ratio.
   The axes are the same as in Figure \ref{fig:growth}.
   $\Mstar=\Msun$, $\Mddot=f\Mgdot$, $t\sub{life}=0.5$ Myr are assumed, and we fixed $\Mgdot=10^{-6}\,\Msunyr$.
   The red solid, green dotted, blue dashed, and violet dot-dashed lines correspond to the dust-to-gas mass ratios, $f=0.01$, $1\fsolar$, $2\fsolar$, and $3\fsolar$, respectively, and $\fsolar$ is the dust-to-gas mass ratio in the solar nebula.
   The black dotted line shows the required initial mass when $f=0.01$ and $\kappa=1$ are assumed, and it corresponds to the plot of equation (\ref{eq:Minit}).
   }
   \label{fig:growth3}
\end{figure}
Figure \ref{fig:growth3} shows the dependence of the initial mass on the dust-to-gas mass ratio, $f$.
The gas accretion rate is fixed as $10^{-6}\,\Msunyr$, and $f$ is treated as a parameter.
Here, we show the cases for $f=0.01$, $1~f\sub{solar}$, $2~f\sub{solar}$, and $3~f\sub{solar}$.

Because the pebble accretion timescale is proportional to $\Mddot^{-1}$, the larger amount of dust lead to faster accretion.
The larger dust-to-gas mass ratio leads to smaller required initial mass, due to the shorter accretion timescale.
When we assume $f=f\sub{solar}$, the required initial mass is $\sim2.0\times10^{-4}\MEarth$, and when $f=3f\sub{solar}$, it falls about $10^{-4}\MEarth$ at 10~au.
However, in the regime when the required initial mass is small, the decrease of the required initial mass becomes small because of the large accretion timescale, which is the same trend in Fig. \ref{fig:growth}.

In Figure \ref{fig:growth3}, the gas accretion rate is fixed to clarify the dependence on the dust-to-gas mass ratio.
However, an increase in the gas accretion rate also changes the accretion of the pebbles significantly.
\begin{figure}
   \centering
   \includegraphics[]{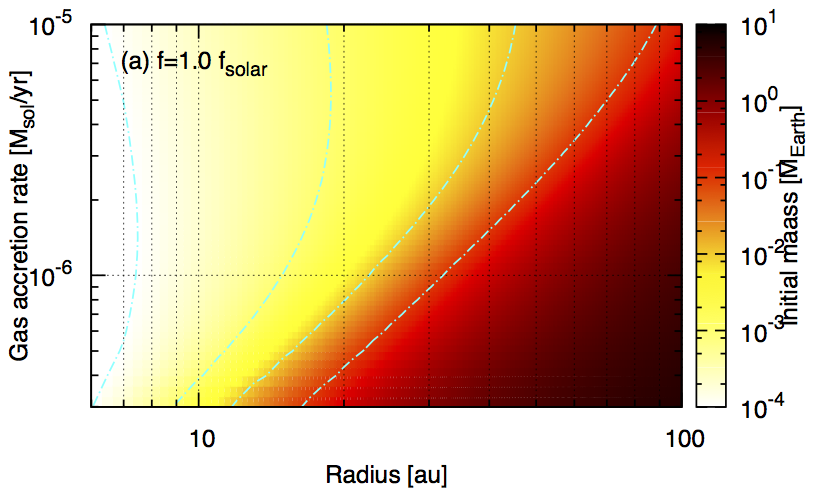}
   \includegraphics[]{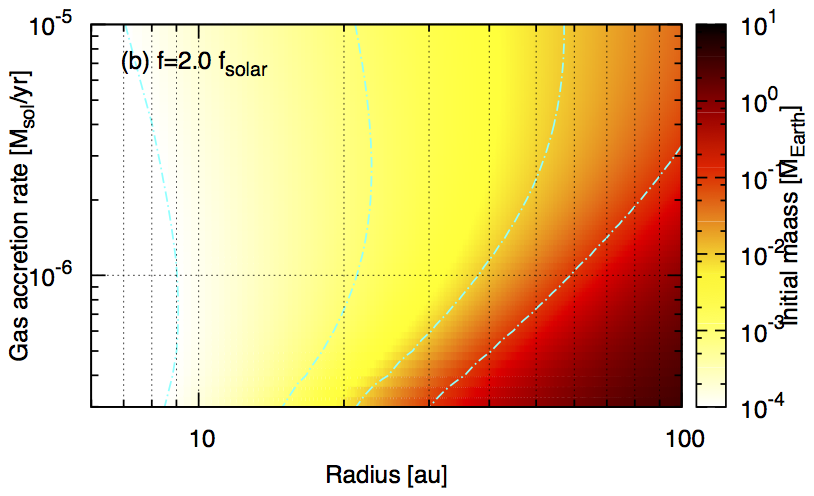}
   \includegraphics[]{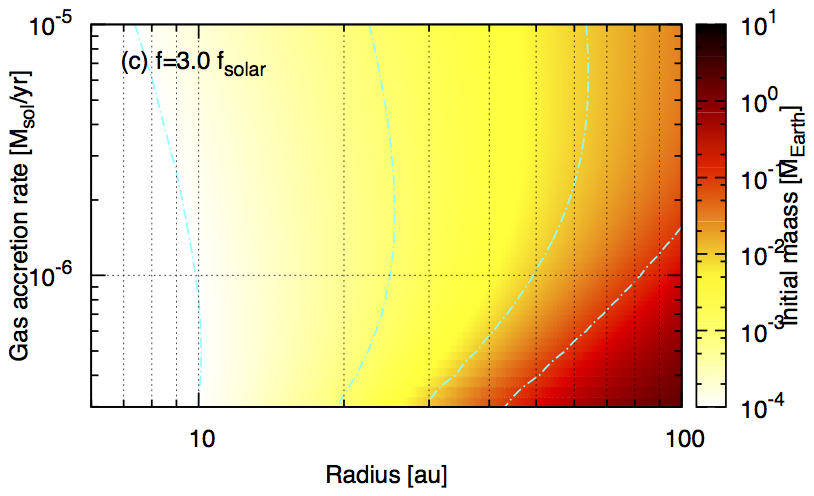}
   \caption{
   Maps of the required initial mass in the radius-gas accretion rate plane.
   The color bars and axes are the same as in Figure \ref{fig:initialmassmap}.
   $\Mstar=\Msun$, $\Mddot=f\Mgdot$, and $t\sub{life}=0.5$ Myr are assumed, and the dust-to-gas mass ratios are (a) $f=1~f\sub{solar}$, (b) $2~f\sub{solar}$, and (c) $3~f\sub{solar}$, respectively.
   The blue dashed lines shows the contour for a required initial mass same as in Fig. \ref{fig:initialmassmap}.
   }
   \label{fig:initialmassmap2}
\end{figure}
Figure \ref{fig:initialmassmap2} shows the maps of the required initial mass in $r-\Mgdot$ planes with different dust-to-gas mass ratios.
The similar dependences of the required initial mass as in Figure \ref{fig:initialmassmap}, are seen in the map.
In the panel (b), for example, the second-leftmost contour shows that the required initial mass decreases as the increase of the gas  accretion rate in the case of $\Mgdot\lesssim 2\times10^{-6}\Msunyr$, but it turns to increase when $\Mgdot\gtrsim 2\times10^{-6}\Msunyr$.

Panel (a) shows the case with the dust-to-gas mass ratio of solar system, which is slightly larger than the case shown in Figure \ref{fig:initialmassmap}.
The required initial mass is smaller in all the region of the map compared to the case of $f=0.01$, due to the large amount of the dust component in the disk.
In this case, for example, the planetary embryo whose mass is $\sim10^{-3}\MEarth$ can grow to 10~$\MEarth$ at $\sim$ 20~au when the gas accretion rate is $\sim4.0\times10^{-6}\Msunyr$.

For more dust rich conditions, as shown in panels (b) and (c), a larger dust component strongly enhances the accretion rate of the pebbles and reduces the required initial mass.
In the case of $f=2\fsolar$, a $\sim10^{-3}\MEarth$ embryo at $\sim20$~au can grow to a 10~$\MEarth$ core when the gas accretion rate is $\sim10^{-6}\Msunyr$.
In the case of $f=3\fsolar$, a $\sim10^{-3}\MEarth$ embryo at $\gtrsim20$~au can grow to a 10~$\MEarth$ core.
In this case, smaller embryo of $\sim10^{-4}\MEarth$ can grow the core at 10~au, even when the gas accretion rate is relatively smaller, $\lesssim10^{-6}\Msunyr$ (shown by the leftmost contour in panel (c)).
This mass is comparable to the mass of Ceres, thus if the planetary embryo whose mass is comparable to Ceres is formed, it can grow to the core of the gas giant.

%%%%%%%%%%%%%%%%%%%%%%%%%%%%%%%%%%%%%%%%%%%%%%%
%%%%%%%%%%%%%%%%%%%%%%%%%%%%%%%%%%%%%%%%%%%%%%%

\subsection{Gas accretion rates and disk lifetimes}\label{results_lifetime}

%%%%%%%%%%%%%%%%%%%%%%%%%%%%%%%%%%%%%%%%%%%%%%%
%%%%%%%%%%%%%%%%%%%%%%%%%%%%%%%%%%%%%%%%%%%%%%%

In the previous sections we assume that the lifetime of the class 0/I object is 0.5 Myr, which is thought to be the typical lifetime of such a disk \citep[e.g.,][]{wc11}.
The lifetime of the disk, however, is not well understood yet.
Therefore it is important to understand the possible effects of the disk lifetime on the formation of planetary cores by the accretion of the pebbles.

\begin{figure}
   \centering
   \includegraphics[]{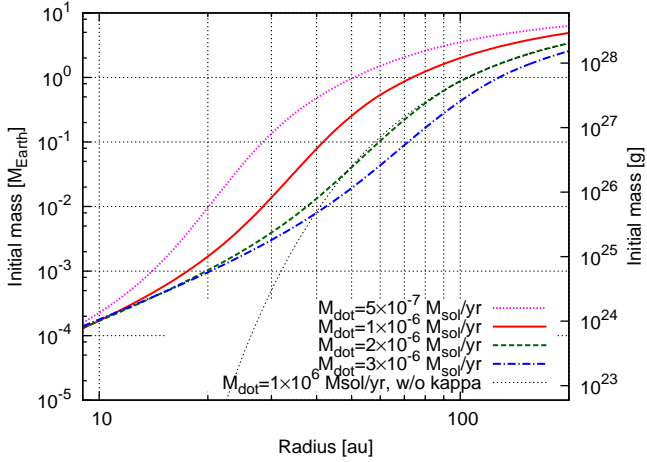} 
   \caption{
   Radial dependence of the required initial mass with different gas accretion rates.
   The meaning of the lines is the same as in Figure \ref{fig:growth}, but the lifetime of the disk is set to $t\sub{life}=1.0$ Myr.
   }
   \label{fig:growth2}
\end{figure}
Figure \ref{fig:growth2} shows the dependence of the required initial mass to form the $10~\MEarth$ cores within the disk lifetime with the different gas accretion rates.
Here we assume that the disk lifetime is $t\sub{life}=1.0$ Myr, twice that in the fiducial model shown in Figure \ref{fig:growth}.
The longer disk lifetime results in a reduced initial mass, compared to Figure \ref{fig:growth}.
In this case, the required initial mass at 10~au is reduced to $\sim2.0\times10^{-4}\MEarth$, and when the gas accretion rate is $2-3\times10^{-6}\Msunyr$, the required initial mass at 20~au is $\sim10^{-3}\MEarth$.
As described above, the accretion timescale is very long when the embryo mass is small, therefore the increase of the gas accretion rate does not contribute to the significant reduction of the required initial mass.
This appears around 10~au in radial distance in Fig. \ref{fig:growth2}.

\begin{figure}
   \centering
   \includegraphics[]{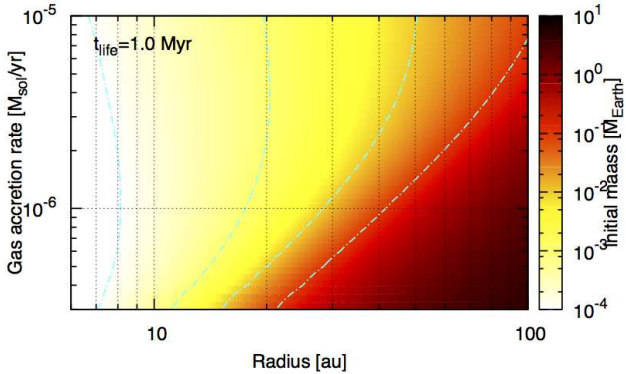} 
   \caption{
   Map of the required initial mass in a radius-gas accretion rate plane.
   The color bar corresponds to the required initial mass to form a 10 $\MEarth$ core.
   $\Mstar=\Msun$, $\Mddot=f\Mgdot$, $t\sub{life}=1.0$ Myr, and $f=0.01$ are assumed in this plot.
   The blue dashed lines shows the contour for a required initial mass same as in Fig. \ref{fig:initialmassmap}.
   }
   \label{fig:initialmassmap3}
\end{figure}
The maps of the required initial mass in $r-\Mgdot$ plane with $t\sub{life}=1.0$~Myr is shown in Fig. \ref{fig:initialmassmap3}.
It is shown that the required initial mass is reduced compared to Fig. \ref{fig:initialmassmap}, because of the longer disk lifetime.

%%%%%%%%%%%%%%%%%%%%%%%%%%%%%%%%%%%%%%%%%%%%%%%
%%%%%%%%%%%%%%%%%%%%%%%%%%%%%%%%%%%%%%%%%%%%%%%

\subsection{Dependence on the stellar mass}\label{results_Mstar}

%%%%%%%%%%%%%%%%%%%%%%%%%%%%%%%%%%%%%%%%%%%%%%%
%%%%%%%%%%%%%%%%%%%%%%%%%%%%%%%%%%%%%%%%%%%%%%%

As shown by equation (\ref{eq:taccthEpstein}), the accretion timescale is a decreasing function of the stellar mass, and therefore core formation by pebble accretion becomes slower around a lower-mass star.
When we consider the dependence of the stellar mass only and ignore the effect from the reduction factor $\kappa$, the accretion timescale is $t\sub{acc}\propto \Mstar^{-3/4}$, and therefore the pebble accretion timescales with $0.5~\Msun$ and $0.3~\Msun$ central stars are $\sim\!1.7$ times and $\sim\!2.5$ times longer than the case for $1.0~\Msun$, respectively.
For the required initial mass, on the other hand, the dependence is not always obvious because it also strongly depends on the disk lifetime; as described in Section \ref{results_lifetime}, the longer disk lifetime is favorable for the formation of the planetary core.

An important indication here is that fast formation of the planet by pebble accretion in the distant region of the disk might be difficult if the stellar mass is relatively small, because of the longer pebble accretion timescale and the small amount of accreting material.
If the dust-to-gas mass ratio is strongly enhanced, the accretion timescale is shorter even in the case of smaller stelar masses, but it is still difficult to form the cores of the giant planets in the distant region in the lifetime of the class 0/I phases.
Therefore, our model suggests that the distant giant planets and/or gaps curved by the planets would hardly be seen around the lower-mass stars, for example, $\lesssim\! 0.5\Msun$, because of the longer pebble accretion timescales.

%%%%%%%%%%%%%%%%%%%%%%%%%%%%%%%%%%%%%%%%%%%%%%%
%%%%%%%%%%%%%%%%%%%%%%%%%%%%%%%%%%%%%%%%%%%%%%%
%%%%%%%%%%%%%%%%%%%%%%%%%%%%%%%%%%%%%%%%%%%%%%%

\section{Discussion}\label{discussion}

%%%%%%%%%%%%%%%%%%%%%%%%%%%%%%%%%%%%%%%%%%%%%%%
%%%%%%%%%%%%%%%%%%%%%%%%%%%%%%%%%%%%%%%%%%%%%%%
%%%%%%%%%%%%%%%%%%%%%%%%%%%%%%%%%%%%%%%%%%%%%%%

%%%%%%%%%%%%%%%%%%%%%%%%%%%%%%%%%%%%%%%%%%%%%%%
%%%%%%%%%%%%%%%%%%%%%%%%%%%%%%%%%%%%%%%%%%%%%%%

\subsection{Comparison with class II disks}\label{discussion_comp}

%%%%%%%%%%%%%%%%%%%%%%%%%%%%%%%%%%%%%%%%%%%%%%%
%%%%%%%%%%%%%%%%%%%%%%%%%%%%%%%%%%%%%%%%%%%%%%%

For class 0/I disks, the relatively larger dust accretion rates supported by the larger gas accretion rate enhance the pebble accretion rate, and it can help the formation of larger planets in the distant orbit in the early stage of star/planet formation.
After the class 0/I phases, a protostar evolves into the class II phase.
The gas and dust accretion rates decrease in this phase, and therefore the accretion timescale will be much longer than that in the class 0/I phases.

\begin{figure}
   \centering
   \includegraphics[]{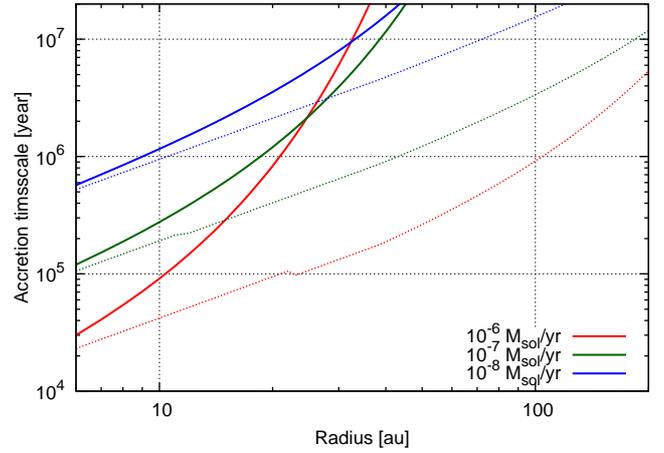} 
   \caption{
   Radial dependence of the accretion timescales.
   The red, green, and blue lines correspond to the difference of the gas accretion rates, $10^{-6}$, $10^{-7}$, and $10^{-8}~\Msunyr$, respectively.
   The line types correspond to the embryo mass; solid and dotted lines correspond to $10^{-3}\MEarth$, $10^{-1}\MEarth$, respectively.
   $\Mddot=f\Mgdot$ and $f=0.01$ are assumed.
   }
   \label{fig:taccretion4}
\end{figure}
Figure \ref{fig:taccretion4} shows the radial dependence of the accretion timescales when the gas accretion rate is significantly reduced.
In this plot we assume that the dust-to-gas mass ratio is $f=0.01$.
The solid and dotted lines correspond to the difference of the embryo mass; the solid lines show the case of $M=10^{-3}\MEarth$, and the dotted lines show the case of $M=10^{-1}\MEarth$.
It is shown that the reduced gas accretion rate significantly prolongs the accretion timescales.
For example, when $\Mgdot=10^{-7}\,\Msunyr$ the accretion timescale exceeds $3.0\times10^{5}$ yr at 10 au, and when $\Mgdot=10^{-8}\,\Msunyr$ it exceeds $3.0\times10^{6}$ yr around 20 au.
This shows that when the gas accretion rate is very small, as during the class II phase, it is very difficult to form the planet or its core in the distant region of the disk within the disk lifetime by pebble accretion, although the disk lifetime itself is about 10 times larger and gas surface density is $10-100$ times smaller.

%%%%%%%%%%%%%%%%%%%%%%%%%%%%%%%%%%%%%%%%%%%%%%%
%%%%%%%%%%%%%%%%%%%%%%%%%%%%%%%%%%%%%%%%%%%%%%%

\subsection{Migration timescales}\label{discussion_migration}

%%%%%%%%%%%%%%%%%%%%%%%%%%%%%%%%%%%%%%%%%%%%%%%
%%%%%%%%%%%%%%%%%%%%%%%%%%%%%%%%%%%%%%%%%%%%%%%

The orbital migration of planets due to disk--planet interaction is one of the important problems in planet formation \citep[e.g.,][]{kn12}.
If the migration of the planet is fast enough, the planet will fall onto the host star unless some mechanisms halt the inward migration.
Therefore it is important to estimate the timescales of the orbital migration.

When the mass of the planet is relatively small and the surface density of the disk is not affected by the planet significantly, type I migration takes place \citep{tan02}.
According to \citet{tan02}, the migration timescale of the type I migration is expressed by
\begin{eqnarray}
\tau\sub{I}=\left(2.7+1.1q\right)^{-1}\frac{\Mstar}{M\sub{p}}\frac{\Mstar}{\Sigma\sub{gas}r^{2}}\frac{\Hgas^{2}}{\Omega},
\end{eqnarray}
where $M\sub{p}$ is the mass of the planet and $q$ is the index of the radial dependence of the gas surface density, $\Sigma\sub{gas}\propto r^{-q}$.
In our disk model, $q=12/7$ as described in equation (\ref{eq:sigmag}), and substituting equations (\ref{eq:aspectratio}), (\ref{eq:Omega}), and (\ref{eq:sigmag}), we obtain
\begin{eqnarray}
\tau\sub{I}\simeq1.1\times10^{5}\left(\frac{M\sub{p}}{\MEarth}\right)^{-1}\tenau^{25/14}\,{\rm yr}.
\end{eqnarray}
Due to the larger surface density of the gas compared to the disk of class II objects, the migration timescale becomes relatively shorter.
Therefore, the planets being formed in class 0/I YSOs tend to quickly accrete onto the central star with this standard formula for type I migration.
However, the migration timescale is sensitive to the mass of the planet and the radial distance.
In addition, the details of the type I migration are still being debated and would depend on the details of the disk \citep{kn12}.

When the mass of the planet is large enough to open a gap in the disk, the orbital migration makes the transition to the type II migration, in which the orbit migrates inward due to the viscous evolution of the disk \citep[e.g.,][]{war97,hi13}.
The upper limit of the migration timescale of the type II migration is
\begin{eqnarray}
\tau\sub{II}=\alpha^{-1}\Hgas^{-2}\Omega^{-1},
\end{eqnarray}
when we assume the migration speed is the same as the radial velocity of the gas.
Substituting the corresponding equations into the above equation, we obtain
\begin{eqnarray}
\tau\sub{II}\simeq6.8\times10^{4}\Mref^{1/2}\Mgdotref^{-1}\nonumber\\
\tenau^{2/7}\,{\rm yr}.
\end{eqnarray}
The timescale of the type II migration is also short, but note that this assumption is the upper limit of the timescale, and the larger mass of the planet slows down the migration rate.

%%%%%%%%%%%%%%%%%%%%%%%%%%%%%%%%%%%%%%%%%%%%%%%
%%%%%%%%%%%%%%%%%%%%%%%%%%%%%%%%%%%%%%%%%%%%%%%

\subsection{Possibility of self-filtration effects}\label{discussion_filtering}

%%%%%%%%%%%%%%%%%%%%%%%%%%%%%%%%%%%%%%%%%%%%%%%
%%%%%%%%%%%%%%%%%%%%%%%%%%%%%%%%%%%%%%%%%%%%%%%

We have shown that core formation by the pebble accretion in the class 0/I disk is efficient even in the distant region, mainly because of the large gas and dust accretion rates.
If the abundance of the dust is enhanced compared to the minimum-mass solar nebula model, the accretion timescales are short, and small objects can grow to form the core of the giant planet around $\sim\!20-30$~au.
Therefore, the cores of the giant planets can form in the early stage of the disk evolution.
However, if the growth of the planetesimals and/or planetary embryos by pebble accretion is too efficient, many embryos accrete the pebbles almost equally, and feeding of the pebbles will decrease.

The filtering efficiency of a single planet is expressed by
\begin{eqnarray}
P=\frac{\dot{M}}{\Mddot}.
\end{eqnarray}
The accretion rate of the pebbles onto the planet is rewritten as $\dot{M}=M/t\sub{acc}$.
In our disk model, the accretion mode in the distant region is the 3D regime as described in Section \ref{2D3D}.
Therefore, by substituting equation (\ref{eq:taccthEpstein}), we obtain the following equation;
\begin{eqnarray}
P\simeq1.0\times10^{-3}\kappa^{2}\Mref^{3/4}\Mgdotref^{-1/4}\nonumber\\
\left(\frac{M}{0.1~\MEarth}\right)\tenau^{-65/56}.
\label{eq:filtering}
\end{eqnarray}
Note that $\kappa^{2}$ will be replaced by $\kappa^{3}$ when the accretion occurs in the Hill regime.
Because the accretion timescale in the 3D regime is inversely proportional to the dust accretion rate, the filtering efficiency does not depend on the dust accretion rate.
Equation (\ref{eq:filtering}) shows that the filtering efficiency in our disk model is $\sim$ 0.1\% at 10 au when the mass of the planet becomes $\sim 0.1~\MEarth$.
Therefore, the effect of self-filtration is insufficient to inhibit the mass growth of the planet except when the mass of the planet becomes very large or when the number of accreting planets is very large.

%%%%%%%%%%%%%%%%%%%%%%%%%%%%%%%%%%%%%%%%%%%%%%%
%%%%%%%%%%%%%%%%%%%%%%%%%%%%%%%%%%%%%%%%%%%%%%%

\subsection{Uncertainty in the dust disk model}\label{uncertainty}

%%%%%%%%%%%%%%%%%%%%%%%%%%%%%%%%%%%%%%%%%%%%%%%
%%%%%%%%%%%%%%%%%%%%%%%%%%%%%%%%%%%%%%%%%%%%%%%

In our model, as described in Section \ref{dustdisk}, we assume perfect sticking for dust evolution.
If we assume icy dust aggregate, perfect sticking may be justified because the relative velocity is $\sim 20~{\rm m\,s^{-1}}$ in our steady state \citep[see Fig. 2 of][]{tsu17}, and it is smaller than the threshold velocity reported by \citet{wad13} of $80~{\rm m\,s^{-1}}$.
Since we focus on the region of the disk outside the water snow line, in this paper, the assumption that dust grains are mainly composed by ice is reasonable.

However, we should note that there is still uncertainty for the threshold velocity.
For example, recent laboratory experiments suggest that the sticking threshold velocity for icy dust aggregates is not as high as \citet{wad13} suggested, $\sim10~{\rm m\,s^{-1}}$ \citep[e.g.,][]{gb15}.

Moreover, composition of dust grains will also affect the sticking threshold.
For example, CO$_{2}$ ice will be important in the outer region of the disk; \citet{mus16} have reported that the sticking threshold velocity for CO$_{2}$ ice is small, and the velocity is reduced even for the mixture of CO$_{2}$ and H$_{2}$O.
Therefore, this uncertainty for the threshold velocity can have an influence on the dust disk model we assumed, and the results.

%%%%%%%%%%%%%%%%%%%%%%%%%%%%%%%%%%%%%%%%%%%%%%%
%%%%%%%%%%%%%%%%%%%%%%%%%%%%%%%%%%%%%%%%%%%%%%%

\subsection{Formation of gas giants during class 0/I phases}\label{formation}

%%%%%%%%%%%%%%%%%%%%%%%%%%%%%%%%%%%%%%%%%%%%%%%
%%%%%%%%%%%%%%%%%%%%%%%%%%%%%%%%%%%%%%%%%%%%%%%

Our model consider the pebble accretion process in the class 0/I disk, and investigate how small mass of the planetesimal/planetary embryo can grow to the core of the giant planet during the lifetime of the class 0/I phases.
In other words, our model gives the initial minimum mass that is necessary to form the gas giants during the lifetime.
In the fiducial case, for example, if a $5.0\times10^{-3}\MEarth$ embryo is formed at 10~au, it can grow up to the $10\MEarth$ core in 0.5~Myr.
In the most optimistic case, i.e., in the dust-rich condition ($f=3\fsolar$), a $10^{-4}\MEarth$ embryo can grow up to the core at 10~au, and $\sim10^{-3}\MEarth$ embryo at $\sim25$~au can grow up to the core.

In our model, the accretion timescale can be short because the dust accretion rate is relatively large, but the estimated values of the required initial mass are not small enough to reach directly by planetesimal formation models.
For example, planetesimals formed via the streaming instability can be $\sim\!10^{-7}~\MEarth$ depending on conditions such as the radial distance \citep[e.g.,][]{joh12}.
Even when the efficient formation processes of planetesimals such as the streaming instability are available, there will be a gap of the mass of several orders of magnitude.
Therefore, additional growth processes should be needed to form the core of the gas giants.

Even if the pebble accretion process during the class 0/I phases are not sufficient to form the giant planets, our model show that the growth of the planetesimal/planetary embryo occurs in the class 0/I phases, thus the embryos with relatively larger masses form at the beginning of the class II phase.
This might alleviate the initial condition of the early formation of the gas giants in the distant orbits in protoplanetary disks.

The formation and growth processes of planetesimals and planetary embryos in the class 0/I phases is not well understood yet.
Therefore, further investigations are needed to clarify whether the cores of the giant planets can be formed during the early phases of the formation of the planetary systems.

%%%%%%%%%%%%%%%%%%%%%%%%%%%%%%%%%%%%%%%%%%%%%%%
%%%%%%%%%%%%%%%%%%%%%%%%%%%%%%%%%%%%%%%%%%%%%%%
%%%%%%%%%%%%%%%%%%%%%%%%%%%%%%%%%%%%%%%%%%%%%%%

\section{Conclusion}\label{conclusion}

%%%%%%%%%%%%%%%%%%%%%%%%%%%%%%%%%%%%%%%%%%%%%%%
%%%%%%%%%%%%%%%%%%%%%%%%%%%%%%%%%%%%%%%%%%%%%%%
%%%%%%%%%%%%%%%%%%%%%%%%%%%%%%%%%%%%%%%%%%%%%%%

We have investigated planet formation by pebble accretion in class 0/I disks, in particular, in the distant region.
We employed an analytical solution of the disk structure of class 0/I objects that was derived by \citet{tsu17}, and estimated the pebble-accretion timescale in the disks on the basis of previous works on pebble accretion in protoplanetary disks \citep{lj12,gui14,ida16}.
Then we give a required mass of a planetesimal/planetary embryo to form a core of the giant planet during the class 0/I phases.

We showed that the required initial mass to form the $10~\MEarth$ core that can grow into the gas giant at 10~au is $\sim\!5.0\times10^{-4}~\MEarth$ when we assume the gas accretion rate is $\sim \! 10^{-6}~\Msunyr$, the dust-to-gas mass ratio is 0.01, and the disk lifetime is 0.5 Myr.
Because the pebble accretion timescale is $\propto \Mgdot^{3/4}$ except for the effect of the reduction factor, the larger gas accretion rate results in a smaller required initial mass, but the longer pebble accretion timescale prevents the initial mass from being quite small values (Figure \ref{fig:growth}).
We also showed that the larger gas accretion rate does not always lead to the smaller pebble accretion timescale when the mass of the planetary embryo is small, because of the reduction effect (Figures \ref{fig:taccretion} and \ref{fig:initialmassmap}).

In addition, the required initial mass is also reduced when the dust-to-gas mass ratio is enhanced.
In the most optimistic case, for example, the formation of the core of the giant planet is possible at $\sim10$~au from the host star when $\Mgdot\sim3.0\times10^{-7}-10^{-6}\,\Msunyr$ and $f=3\fsolar$, and the embryo mass is comparable to Ceres's mass, $\sim10^{-4}\MEarth$ (Figure \ref{fig:initialmassmap2}(c)).

We also considered the effect of the lifetime of the class 0/I phases.
The required initial mass is reduced when the disk lifetime is longer, because its dependence is $\propto\exp\!\left(-t\sub{life}/t\sub{acc}\right)$ when the embryo mass is larger.
If the lifetime of the class 0/I phases are 1.0 Myr, the required initial mass at 10~au is reduced to $\sim2.0\times10^{-4}\MEarth$, and when the gas accretion rate is $2-3\times10^{-6}\Msunyr$, the required initial mass at 20~au is $\sim10^{-3}\MEarth$ (Figures \ref{fig:growth2} and \ref{fig:initialmassmap3}).
Therefore, the longer lifetime of the class 0/I phases is favored for the formation of larger planets in the distant region of the disk.

Our model shows the pebble accretion timescale in the class 0/I phases and gives the required mass of planetary embryos to form the cores of the giant planets.
If large bodies whose mass is comparable to Ceres exist, they can grow to $10\MEarth$ cores during the lifetime of the class 0/I phases.
It is unclear that such large objects can form in the disk, and more detailed works will be needed to investigate whether distant gas giants which explain the gaps and rings in the young protoplanetary disks can be formed or not.
In any case, our model suggests that planet formation in the early stage of the star/planet formation processes is important.

\section*{Acknowledgements}

We thank the anonymous referee for their useful reports that greatly improved the manuscript.
This work was supported by MEXT Leading Initiative for Excellent Young Researchers and JSPS KAKENHI Grant Number JP18K13581.

%%%%%%%%%%%%%%%%%%%%%%%%%%%%%%%%%%%%%%%%%%%%%%%%%%



\bibliographystyle{mnras}
\bibliography{ref} % if your bibtex file is called example.bib

\begin{thebibliography}{}
\makeatletter
\relax
\def\mn@urlcharsother{\let\do\@makeother \do\$\do\&\do\#\do\^\do\_\do\%\do\~}
\def\mn@doi{\begingroup\mn@urlcharsother \@ifnextchar [ {\mn@doi@}
  {\mn@doi@[]}}
\def\mn@doi@[#1]#2{\def\@tempa{#1}\ifx\@tempa\@empty \href
  {http://dx.doi.org/#2} {doi:#2}\else \href {http://dx.doi.org/#2} {#1}\fi
  \endgroup}
\def\mn@eprint#1#2{\mn@eprint@#1:#2::\@nil}
\def\mn@eprint@arXiv#1{\href {http://arxiv.org/abs/#1} {{\tt arXiv:#1}}}
\def\mn@eprint@dblp#1{\href {http://dblp.uni-trier.de/rec/bibtex/#1.xml}
  {dblp:#1}}
\def\mn@eprint@#1:#2:#3:#4\@nil{\def\@tempa {#1}\def\@tempb {#2}\def\@tempc
  {#3}\ifx \@tempc \@empty \let \@tempc \@tempb \let \@tempb \@tempa \fi \ifx
  \@tempb \@empty \def\@tempb {arXiv}\fi \@ifundefined
  {mn@eprint@\@tempb}{\@tempb:\@tempc}{\expandafter \expandafter \csname
  mn@eprint@\@tempb\endcsname \expandafter{\@tempc}}}

\bibitem[\protect\citeauthoryear{{ALMA Partnership} et~al.,}{{ALMA Partnership}
  et~al.}{2015}]{alm15}
{ALMA Partnership} et~al., 2015, \mn@doi [\apjl] {10.1088/2041-8205/808/1/L3},
  \href {http://adsabs.harvard.edu/abs/2015ApJ...808L...3A} {808, L3}

\bibitem[\protect\citeauthoryear{{Akiyama}, {Hasegawa}, {Hayashi}  \&
  {Iguchi}}{{Akiyama} et~al.}{2016}]{aki16}
{Akiyama} E.,  {Hasegawa} Y.,  {Hayashi} M.,   {Iguchi} S.,  2016, \mn@doi
  [\apj] {10.3847/0004-637X/818/2/158}, \href
  {http://adsabs.harvard.edu/abs/2016ApJ...818..158A} {818, 158}

\bibitem[\protect\citeauthoryear{{Andrews}, {Wilner}, {Hughes}, {Qi}  \&
  {Dullemond}}{{Andrews} et~al.}{2010}]{and10}
{Andrews} S.~M.,  {Wilner} D.~J.,  {Hughes} A.~M.,  {Qi} C.,   {Dullemond}
  C.~P.,  2010, \mn@doi [\apj] {10.1088/0004-637X/723/2/1241}, \href
  {http://adsabs.harvard.edu/abs/2010ApJ...723.1241A} {723, 1241}

\bibitem[\protect\citeauthoryear{{Andrews} et~al.,}{{Andrews}
  et~al.}{2016}]{and16}
{Andrews} S.~M.,  et~al., 2016, \mn@doi [\apjl] {10.3847/2041-8205/820/2/L40},
  \href {http://adsabs.harvard.edu/abs/2016ApJ...820L..40A} {820, L40}

\bibitem[\protect\citeauthoryear{{Ansdell} et~al.,}{{Ansdell}
  et~al.}{2016}]{ans16}
{Ansdell} M.,  et~al., 2016, \mn@doi [\apj] {10.3847/0004-637X/828/1/46}, \href
  {http://adsabs.harvard.edu/abs/2016ApJ...828...46A} {828, 46}

\bibitem[\protect\citeauthoryear{{Carrera}, {Gorti}, {Johansen}  \&
  {Davies}}{{Carrera} et~al.}{2017}]{car17}
{Carrera} D.,  {Gorti} U.,  {Johansen} A.,   {Davies} M.~B.,  2017, \mn@doi
  [\apj] {10.3847/1538-4357/aa6932}, \href
  {http://adsabs.harvard.edu/abs/2017ApJ...839...16C} {839, 16}

\bibitem[\protect\citeauthoryear{{Chatterjee} \& {Tan}}{{Chatterjee} \&
  {Tan}}{2014}]{ct14}
{Chatterjee} S.,  {Tan} J.~C.,  2014, \mn@doi [\apj]
  {10.1088/0004-637X/780/1/53}, \href
  {http://adsabs.harvard.edu/abs/2014ApJ...780...53C} {780, 53}

\bibitem[\protect\citeauthoryear{{Chiang} \& {Goldreich}}{{Chiang} \&
  {Goldreich}}{1997}]{cg97}
{Chiang} E.~I.,  {Goldreich} P.,  1997, \mn@doi [\apj] {10.1086/304869}, \href
  {http://adsabs.harvard.edu/abs/1997ApJ...490..368C} {490, 368}

\bibitem[\protect\citeauthoryear{{Cieza} et~al.,}{{Cieza} et~al.}{2017}]{cie17}
{Cieza} L.~A.,  et~al., 2017, \mn@doi [\apjl] {10.3847/2041-8213/aa9b7b}, \href
  {http://adsabs.harvard.edu/abs/2017ApJ...851L..23C} {851, L23}

\bibitem[\protect\citeauthoryear{{Dipierro}, {Price}, {Laibe}, {Hirsh},
  {Cerioli}  \& {Lodato}}{{Dipierro} et~al.}{2015}]{dip15}
{Dipierro} G.,  {Price} D.,  {Laibe} G.,  {Hirsh} K.,  {Cerioli} A.,   {Lodato}
  G.,  2015, \mn@doi [\mnras] {10.1093/mnrasl/slv105}, \href
  {http://adsabs.harvard.edu/abs/2015MNRAS.453L..73D} {453, L73}

\bibitem[\protect\citeauthoryear{{Dodson-Robinson}, {Veras}, {Ford}  \&
  {Beichman}}{{Dodson-Robinson} et~al.}{2009}]{dod09}
{Dodson-Robinson} S.~E.,  {Veras} D.,  {Ford} E.~B.,   {Beichman} C.~A.,  2009,
  \mn@doi [\apj] {10.1088/0004-637X/707/1/79}, \href
  {http://adsabs.harvard.edu/abs/2009ApJ...707...79D} {707, 79}

\bibitem[\protect\citeauthoryear{{Dong}, {Zhu}  \& {Whitney}}{{Dong}
  et~al.}{2015}]{don15}
{Dong} R.,  {Zhu} Z.,   {Whitney} B.,  2015, \mn@doi [\apj]
  {10.1088/0004-637X/809/1/93}, \href
  {http://adsabs.harvard.edu/abs/2015ApJ...809...93D} {809, 93}

\bibitem[\protect\citeauthoryear{{Dzyurkevich}, {Flock}, {Turner}, {Klahr}  \&
  {Henning}}{{Dzyurkevich} et~al.}{2010}]{dzy10}
{Dzyurkevich} N.,  {Flock} M.,  {Turner} N.~J.,  {Klahr} H.,   {Henning} T.,
  2010, \mn@doi [\aap] {10.1051/0004-6361/200912834}, \href
  {http://adsabs.harvard.edu/abs/2010A%26A...515A..70D} {515, A70}

\bibitem[\protect\citeauthoryear{{Ercolano}, {Rosotti}, {Picogna}  \&
  {Testi}}{{Ercolano} et~al.}{2017}]{erc17}
{Ercolano} B.,  {Rosotti} G.~P.,  {Picogna} G.,   {Testi} L.,  2017, \mn@doi
  [\mnras] {10.1093/mnrasl/slw188}, \href
  {http://adsabs.harvard.edu/abs/2017MNRAS.464L..95E} {464, L95}

\bibitem[\protect\citeauthoryear{{Fedele} et~al.,}{{Fedele}
  et~al.}{2018}]{fed18}
{Fedele} D.,  et~al., 2018, \mn@doi [\aap] {10.1051/0004-6361/201731978}, \href
  {http://adsabs.harvard.edu/abs/2018A%26A...610A..24F} {610, A24}

\bibitem[\protect\citeauthoryear{{Flock}, {Ruge}, {Dzyurkevich}, {Henning},
  {Klahr}  \& {Wolf}}{{Flock} et~al.}{2015}]{flo15}
{Flock} M.,  {Ruge} J.~P.,  {Dzyurkevich} N.,  {Henning} T.,  {Klahr} H.,
  {Wolf} S.,  2015, \mn@doi [\aap] {10.1051/0004-6361/201424693}, \href
  {http://adsabs.harvard.edu/abs/2015A%26A...574A..68F} {574, A68}

\bibitem[\protect\citeauthoryear{{Goldreich} \& {Tremaine}}{{Goldreich} \&
  {Tremaine}}{1980}]{gt80}
{Goldreich} P.,  {Tremaine} S.,  1980, \mn@doi [\apj] {10.1086/158356}, \href
  {http://adsabs.harvard.edu/abs/1980ApJ...241..425G} {241, 425}

\bibitem[\protect\citeauthoryear{{Guillot}, {Ida}  \& {Ormel}}{{Guillot}
  et~al.}{2014}]{gui14}
{Guillot} T.,  {Ida} S.,   {Ormel} C.~W.,  2014, \mn@doi [\aap]
  {10.1051/0004-6361/201323021}, \href
  {http://adsabs.harvard.edu/abs/2014A%26A...572A..72G} {572, A72}

\bibitem[\protect\citeauthoryear{{Gundlach} \& {Blum}}{{Gundlach} \&
  {Blum}}{2015}]{gb15}
{Gundlach} B.,  {Blum} J.,  2015, \mn@doi [\apj] {10.1088/0004-637X/798/1/34},
  \href {http://adsabs.harvard.edu/abs/2015ApJ...798...34G} {798, 34}

\bibitem[\protect\citeauthoryear{{Hasegawa} \& {Ida}}{{Hasegawa} \&
  {Ida}}{2013}]{hi13}
{Hasegawa} Y.,  {Ida} S.,  2013, \mn@doi [\apj] {10.1088/0004-637X/774/2/146},
  \href {http://adsabs.harvard.edu/abs/2013ApJ...774..146H} {774, 146}

\bibitem[\protect\citeauthoryear{{Hayashi}, {Nakazawa}  \&
  {Nakagawa}}{{Hayashi} et~al.}{1985}]{hay85}
{Hayashi} C.,  {Nakazawa} K.,   {Nakagawa} Y.,  1985, in {Black} D.~C.,
  {Matthews} M.~S.,  eds, Protostars and Planets II. pp 1100--1153

\bibitem[\protect\citeauthoryear{{Hueso} \& {Guillot}}{{Hueso} \&
  {Guillot}}{2005}]{hg05}
{Hueso} R.,  {Guillot} T.,  2005, \mn@doi [\aap] {10.1051/0004-6361:20041905},
  \href {http://ads.nao.ac.jp/abs/2005A%26A...442..703H} {442, 703}

\bibitem[\protect\citeauthoryear{{Hunter}}{{Hunter}}{1977}]{hun77}
{Hunter} C.,  1977, \mn@doi [\apj] {10.1086/155739}, \href
  {http://adsabs.harvard.edu/abs/1977ApJ...218..834H} {218, 834}

\bibitem[\protect\citeauthoryear{{Ida}, {Guillot}  \& {Morbidelli}}{{Ida}
  et~al.}{2016}]{ida16}
{Ida} S.,  {Guillot} T.,   {Morbidelli} A.,  2016, \mn@doi [\aap]
  {10.1051/0004-6361/201628099}, \href
  {http://adsabs.harvard.edu/abs/2016A%26A...591A..72I} {591, A72}

\bibitem[\protect\citeauthoryear{{Ikoma}, {Nakazawa}  \& {Emori}}{{Ikoma}
  et~al.}{2000}]{iko00}
{Ikoma} M.,  {Nakazawa} K.,   {Emori} H.,  2000, \mn@doi [\apj]
  {10.1086/309050}, \href {http://adsabs.harvard.edu/abs/2000ApJ...537.1013I}
  {537, 1013}

\bibitem[\protect\citeauthoryear{{Isella} et~al.,}{{Isella}
  et~al.}{2016}]{ise16}
{Isella} A.,  et~al., 2016, \mn@doi [Physical Review Letters]
  {10.1103/PhysRevLett.117.251101}, \href
  {http://adsabs.harvard.edu/abs/2016PhRvL.117y1101I} {117, 251101}

\bibitem[\protect\citeauthoryear{{Jin}, {Li}, {Isella}, {Li}  \& {Ji}}{{Jin}
  et~al.}{2016}]{jin16}
{Jin} S.,  {Li} S.,  {Isella} A.,  {Li} H.,   {Ji} J.,  2016, \mn@doi [\apj]
  {10.3847/0004-637X/818/1/76}, \href
  {http://adsabs.harvard.edu/abs/2016ApJ...818...76J} {818, 76}

\bibitem[\protect\citeauthoryear{{Johansen}, {Oishi}, {Mac Low}, {Klahr},
  {Henning}  \& {Youdin}}{{Johansen} et~al.}{2007}]{joh07}
{Johansen} A.,  {Oishi} J.~S.,  {Mac Low} M.-M.,  {Klahr} H.,  {Henning} T.,
  {Youdin} A.,  2007, \mn@doi [\nat] {10.1038/nature06086}, \href
  {http://adsabs.harvard.edu/abs/2007Natur.448.1022J} {448, 1022}

\bibitem[\protect\citeauthoryear{{Johansen}, {Youdin}  \& {Klahr}}{{Johansen}
  et~al.}{2009}]{joh09}
{Johansen} A.,  {Youdin} A.,   {Klahr} H.,  2009, \mn@doi [\apj]
  {10.1088/0004-637X/697/2/1269}, \href
  {http://adsabs.harvard.edu/abs/2009ApJ...697.1269J} {697, 1269}

\bibitem[\protect\citeauthoryear{{Johansen}, {Youdin}  \&
  {Lithwick}}{{Johansen} et~al.}{2012}]{joh12}
{Johansen} A.,  {Youdin} A.~N.,   {Lithwick} Y.,  2012, \mn@doi [\aap]
  {10.1051/0004-6361/201117701}, \href
  {http://adsabs.harvard.edu/abs/2012A%26A...537A.125J} {537, A125}

\bibitem[\protect\citeauthoryear{{Johansen}, {Mac Low}, {Lacerda}  \&
  {Bizzarro}}{{Johansen} et~al.}{2015}]{joh15}
{Johansen} A.,  {Mac Low} M.-M.,  {Lacerda} P.,   {Bizzarro} M.,  2015, \mn@doi
  [Science Advances] {10.1126/sciadv.1500109}, \href
  {http://adsabs.harvard.edu/abs/2015SciA....1E0109J} {1, 1500109}

\bibitem[\protect\citeauthoryear{{Kanagawa}, {Muto}, {Tanaka}, {Tanigawa},
  {Takeuchi}, {Tsukagoshi}  \& {Momose}}{{Kanagawa} et~al.}{2015}]{kan15}
{Kanagawa} K.~D.,  {Muto} T.,  {Tanaka} H.,  {Tanigawa} T.,  {Takeuchi} T.,
  {Tsukagoshi} T.,   {Momose} M.,  2015, \mn@doi [\apjl]
  {10.1088/2041-8205/806/1/L15}, \href
  {http://adsabs.harvard.edu/abs/2015ApJ...806L..15K} {806, L15}

\bibitem[\protect\citeauthoryear{{Kanagawa}, {Muto}, {Tanaka}, {Tanigawa},
  {Takeuchi}, {Tsukagoshi}  \& {Momose}}{{Kanagawa} et~al.}{2016}]{kan16}
{Kanagawa} K.~D.,  {Muto} T.,  {Tanaka} H.,  {Tanigawa} T.,  {Takeuchi} T.,
  {Tsukagoshi} T.,   {Momose} M.,  2016, \mn@doi [\pasj] {10.1093/pasj/psw037},
  \href {http://adsabs.harvard.edu/abs/2016PASJ...68...43K} {68, 43}

\bibitem[\protect\citeauthoryear{{Kataoka}, {Tanaka}, {Okuzumi}  \&
  {Wada}}{{Kataoka} et~al.}{2013}]{kat13}
{Kataoka} A.,  {Tanaka} H.,  {Okuzumi} S.,   {Wada} K.,  2013, \mn@doi [\aap]
  {10.1051/0004-6361/201322151}, \href
  {http://adsabs.harvard.edu/abs/2013A%26A...557L...4K} {557, L4}

\bibitem[\protect\citeauthoryear{{Kley} \& {Nelson}}{{Kley} \&
  {Nelson}}{2012}]{kn12}
{Kley} W.,  {Nelson} R.~P.,  2012, \mn@doi [\araa]
  {10.1146/annurev-astro-081811-125523}, \href
  {http://adsabs.harvard.edu/abs/2012ARA%26A..50..211K} {50, 211}

\bibitem[\protect\citeauthoryear{{Kokubo} \& {Ida}}{{Kokubo} \&
  {Ida}}{1998}]{ki98}
{Kokubo} E.,  {Ida} S.,  1998, \mn@doi [\icarus] {10.1006/icar.1997.5840},
  \href {http://adsabs.harvard.edu/abs/1998Icar..131..171K} {131, 171}

\bibitem[\protect\citeauthoryear{{Kratter}, {Murray-Clay}  \&
  {Youdin}}{{Kratter} et~al.}{2010}]{kra10}
{Kratter} K.~M.,  {Murray-Clay} R.~A.,   {Youdin} A.~N.,  2010, \mn@doi [\apj]
  {10.1088/0004-637X/710/2/1375}, \href
  {http://adsabs.harvard.edu/abs/2010ApJ...710.1375K} {710, 1375}

\bibitem[\protect\citeauthoryear{{Kretke} \& {Lin}}{{Kretke} \&
  {Lin}}{2007}]{kl07}
{Kretke} K.~A.,  {Lin} D.~N.~C.,  2007, \mn@doi [\apjl] {10.1086/520718}, \href
  {http://adsabs.harvard.edu/abs/2007ApJ...664L..55K} {664, L55}

\bibitem[\protect\citeauthoryear{{Kusaka}, {Nakano}  \& {Hayashi}}{{Kusaka}
  et~al.}{1970}]{kus70}
{Kusaka} T.,  {Nakano} T.,   {Hayashi} C.,  1970, \mn@doi [Progress of
  Theoretical Physics] {10.1143/PTP.44.1580}, \href
  {http://adsabs.harvard.edu/abs/1970PThPh..44.1580K} {44, 1580}

\bibitem[\protect\citeauthoryear{{Lambrechts} \& {Johansen}}{{Lambrechts} \&
  {Johansen}}{2012}]{lj12}
{Lambrechts} M.,  {Johansen} A.,  2012, \mn@doi [\aap]
  {10.1051/0004-6361/201219127}, \href
  {http://adsabs.harvard.edu/abs/2012A%26A...544A..32L} {544, A32}

\bibitem[\protect\citeauthoryear{{Lambrechts} \& {Johansen}}{{Lambrechts} \&
  {Johansen}}{2014}]{lj14}
{Lambrechts} M.,  {Johansen} A.,  2014, \mn@doi [\aap]
  {10.1051/0004-6361/201424343}, \href
  {http://adsabs.harvard.edu/abs/2014A%26A...572A.107L} {572, A107}

\bibitem[\protect\citeauthoryear{{Lambrechts}, {Johansen}  \&
  {Morbidelli}}{{Lambrechts} et~al.}{2014}]{lam14}
{Lambrechts} M.,  {Johansen} A.,   {Morbidelli} A.,  2014, \mn@doi [\aap]
  {10.1051/0004-6361/201423814}, \href
  {http://adsabs.harvard.edu/abs/2014A%26A...572A..35L} {572, A35}

\bibitem[\protect\citeauthoryear{{Larson}}{{Larson}}{1969}]{lar69}
{Larson} R.~B.,  1969, \mn@doi [\mnras] {10.1093/mnras/145.3.271}, \href
  {http://adsabs.harvard.edu/abs/1969MNRAS.145..271L} {145, 271}

\bibitem[\protect\citeauthoryear{{Lin} \& {Papaloizou}}{{Lin} \&
  {Papaloizou}}{1986}]{lp86}
{Lin} D.~N.~C.,  {Papaloizou} J.,  1986, \mn@doi [\apj] {10.1086/164426}, \href
  {http://adsabs.harvard.edu/abs/1986ApJ...307..395L} {307, 395}

\bibitem[\protect\citeauthoryear{{Liseau} et~al.,}{{Liseau}
  et~al.}{2015}]{lis15}
{Liseau} R.,  et~al., 2015, \mn@doi [\aap] {10.1051/0004-6361/201525641}, \href
  {http://adsabs.harvard.edu/abs/2015A%26A...578A.131L} {578, A131}

\bibitem[\protect\citeauthoryear{{Lodders}}{{Lodders}}{2003}]{lod03}
{Lodders} K.,  2003, \mn@doi [\apj] {10.1086/375492}, \href
  {http://adsabs.harvard.edu/abs/2003ApJ...591.1220L} {591, 1220}

\bibitem[\protect\citeauthoryear{{Loomis}, {{\"O}berg}, {Andrews}  \&
  {MacGregor}}{{Loomis} et~al.}{2017}]{loo17}
{Loomis} R.~A.,  {{\"O}berg} K.~I.,  {Andrews} S.~M.,   {MacGregor} M.~A.,
  2017, \mn@doi [\apj] {10.3847/1538-4357/aa6c63}, \href
  {http://adsabs.harvard.edu/abs/2017ApJ...840...23L} {840, 23}

\bibitem[\protect\citeauthoryear{{Machida}, {Inutsuka}  \&
  {Matsumoto}}{{Machida} et~al.}{2011}]{mac11}
{Machida} M.~N.,  {Inutsuka} S.-I.,   {Matsumoto} T.,  2011, \mn@doi [\pasj]
  {10.1093/pasj/63.3.555}, \href
  {http://adsabs.harvard.edu/abs/2011PASJ...63..555M} {63, 555}

\bibitem[\protect\citeauthoryear{{Masson}, {Chabrier}, {Hennebelle}, {Vaytet}
  \& {Commer{\c c}on}}{{Masson} et~al.}{2016}]{mas16}
{Masson} J.,  {Chabrier} G.,  {Hennebelle} P.,  {Vaytet} N.,   {Commer{\c c}on}
  B.,  2016, \mn@doi [\aap] {10.1051/0004-6361/201526371}, \href
  {http://adsabs.harvard.edu/abs/2016A%26A...587A..32M} {587, A32}

\bibitem[\protect\citeauthoryear{{Mizuno}}{{Mizuno}}{1980}]{miz80}
{Mizuno} H.,  1980, \mn@doi [Progress of Theoretical Physics]
  {10.1143/PTP.64.544}, \href
  {http://adsabs.harvard.edu/abs/1980PThPh..64..544M} {64, 544}

\bibitem[\protect\citeauthoryear{{Morbidelli} et~al.,}{{Morbidelli}
  et~al.}{2016}]{mor16}
{Morbidelli} A.,  et~al., 2016, \mn@doi [\icarus]
  {10.1016/j.icarus.2015.11.027}, \href
  {http://adsabs.harvard.edu/abs/2016Icar..267..368M} {267, 368}

\bibitem[\protect\citeauthoryear{{Musiolik}, {Teiser}, {Jankowski}  \&
  {Wurm}}{{Musiolik} et~al.}{2016}]{mus16}
{Musiolik} G.,  {Teiser} J.,  {Jankowski} T.,   {Wurm} G.,  2016, \mn@doi
  [\apj] {10.3847/0004-637X/827/1/63}, \href
  {http://adsabs.harvard.edu/abs/2016ApJ...827...63M} {827, 63}

\bibitem[\protect\citeauthoryear{{Nakamoto} \& {Nakagawa}}{{Nakamoto} \&
  {Nakagawa}}{1994}]{nn94}
{Nakamoto} T.,  {Nakagawa} Y.,  1994, \mn@doi [\apj] {10.1086/173678}, \href
  {http://adsabs.harvard.edu/abs/1994ApJ...421..640N} {421, 640}

\bibitem[\protect\citeauthoryear{{Okuzumi}, {Tanaka}, {Kobayashi}  \&
  {Wada}}{{Okuzumi} et~al.}{2012}]{oku12}
{Okuzumi} S.,  {Tanaka} H.,  {Kobayashi} H.,   {Wada} K.,  2012, \mn@doi [\apj]
  {10.1088/0004-637X/752/2/106}, \href
  {http://adsabs.harvard.edu/abs/2012ApJ...752..106O} {752, 106}

\bibitem[\protect\citeauthoryear{{Okuzumi}, {Momose}, {Sirono}, {Kobayashi}  \&
  {Tanaka}}{{Okuzumi} et~al.}{2016}]{oku16}
{Okuzumi} S.,  {Momose} M.,  {Sirono} S.-i.,  {Kobayashi} H.,   {Tanaka} H.,
  2016, \mn@doi [\apj] {10.3847/0004-637X/821/2/82}, \href
  {http://adsabs.harvard.edu/abs/2016ApJ...821...82O} {821, 82}

\bibitem[\protect\citeauthoryear{{Ormel} \& {Cuzzi}}{{Ormel} \&
  {Cuzzi}}{2007}]{oc07}
{Ormel} C.~W.,  {Cuzzi} J.~N.,  2007, \mn@doi [\aap]
  {10.1051/0004-6361:20066899}, \href
  {http://adsabs.harvard.edu/abs/2007A%26A...466..413O} {466, 413}

\bibitem[\protect\citeauthoryear{{Ormel} \& {Klahr}}{{Ormel} \&
  {Klahr}}{2010}]{ok10}
{Ormel} C.~W.,  {Klahr} H.~H.,  2010, \mn@doi [\aap]
  {10.1051/0004-6361/201014903}, \href
  {http://adsabs.harvard.edu/abs/2010A%26A...520A..43O} {520, A43}

\bibitem[\protect\citeauthoryear{{Ormel} \& {Kobayashi}}{{Ormel} \&
  {Kobayashi}}{2012}]{ok12}
{Ormel} C.~W.,  {Kobayashi} H.,  2012, \mn@doi [\apj]
  {10.1088/0004-637X/747/2/115}, \href
  {http://adsabs.harvard.edu/abs/2012ApJ...747..115O} {747, 115}

\bibitem[\protect\citeauthoryear{{Penston}}{{Penston}}{1969}]{pen69}
{Penston} M.~V.,  1969, \mn@doi [\mnras] {10.1093/mnras/144.4.425}, \href
  {http://adsabs.harvard.edu/abs/1969MNRAS.144..425P} {144, 425}

\bibitem[\protect\citeauthoryear{{Picogna} \& {Kley}}{{Picogna} \&
  {Kley}}{2015}]{pk15}
{Picogna} G.,  {Kley} W.,  2015, \mn@doi [\aap] {10.1051/0004-6361/201526921},
  \href {http://adsabs.harvard.edu/abs/2015A%26A...584A.110P} {584, A110}

\bibitem[\protect\citeauthoryear{{Pollack}, {Hubickyj}, {Bodenheimer},
  {Lissauer}, {Podolak}  \& {Greenzweig}}{{Pollack} et~al.}{1996}]{pol96}
{Pollack} J.~B.,  {Hubickyj} O.,  {Bodenheimer} P.,  {Lissauer} J.~J.,
  {Podolak} M.,   {Greenzweig} Y.,  1996, \mn@doi [\icarus]
  {10.1006/icar.1996.0190}, \href
  {http://adsabs.harvard.edu/abs/1996Icar..124...62P} {124, 62}

\bibitem[\protect\citeauthoryear{{Rafikov}}{{Rafikov}}{2011}]{raf11}
{Rafikov} R.~R.,  2011, \mn@doi [\apj] {10.1088/0004-637X/727/2/86}, \href
  {http://adsabs.harvard.edu/abs/2011ApJ...727...86R} {727, 86}

\bibitem[\protect\citeauthoryear{{Safronov}}{{Safronov}}{1972}]{saf72}
{Safronov} V.~S.,  1972, {Evolution of the protoplanetary cloud and formation
  of the earth and planets.}

\bibitem[\protect\citeauthoryear{{Sato}, {Okuzumi}  \& {Ida}}{{Sato}
  et~al.}{2016}]{sat16}
{Sato} T.,  {Okuzumi} S.,   {Ida} S.,  2016, \mn@doi [\aap]
  {10.1051/0004-6361/201527069}, \href
  {http://adsabs.harvard.edu/abs/2016A%26A...589A..15S} {589, A15}

\bibitem[\protect\citeauthoryear{{Shakura} \& {Sunyaev}}{{Shakura} \&
  {Sunyaev}}{1973}]{ss73}
{Shakura} N.~I.,  {Sunyaev} R.~A.,  1973, \aap, \href
  {http://adsabs.harvard.edu/abs/1973A%26A....24..337S} {24, 337}

\bibitem[\protect\citeauthoryear{{Shu}}{{Shu}}{1977}]{shu77}
{Shu} F.~H.,  1977, \mn@doi [\apj] {10.1086/155274}, \href
  {http://adsabs.harvard.edu/abs/1977ApJ...214..488S} {214, 488}

\bibitem[\protect\citeauthoryear{{Siess}, {Dufour}  \& {Forestini}}{{Siess}
  et~al.}{2000}]{sie00}
{Siess} L.,  {Dufour} E.,   {Forestini} M.,  2000, \aap, \href
  {http://adsabs.harvard.edu/abs/2000A%26A...358..593S} {358, 593}

\bibitem[\protect\citeauthoryear{{Stamatellos} \& {Whitworth}}{{Stamatellos} \&
  {Whitworth}}{2008}]{sw08}
{Stamatellos} D.,  {Whitworth} A.~P.,  2008, \mn@doi [\aap]
  {10.1051/0004-6361:20078628}, \href
  {http://adsabs.harvard.edu/abs/2008A%26A...480..879S} {480, 879}

\bibitem[\protect\citeauthoryear{{Stamatellos} \& {Whitworth}}{{Stamatellos} \&
  {Whitworth}}{2009}]{sw09}
{Stamatellos} D.,  {Whitworth} A.~P.,  2009, \mn@doi [\mnras]
  {10.1111/j.1365-2966.2008.14069.x}, \href
  {http://adsabs.harvard.edu/abs/2009MNRAS.392..413S} {392, 413}

\bibitem[\protect\citeauthoryear{{Stevenson}}{{Stevenson}}{1982}]{ste82}
{Stevenson} D.~J.,  1982, \mn@doi [\planss] {10.1016/0032-0633(82)90108-8},
  \href {http://adsabs.harvard.edu/abs/1982P%26SS...30..755S} {30, 755}

\bibitem[\protect\citeauthoryear{{Takahashi} \& {Inutsuka}}{{Takahashi} \&
  {Inutsuka}}{2014}]{ti14}
{Takahashi} S.~Z.,  {Inutsuka} S.-i.,  2014, \mn@doi [\apj]
  {10.1088/0004-637X/794/1/55}, \href
  {http://adsabs.harvard.edu/abs/2014ApJ...794...55T} {794, 55}

\bibitem[\protect\citeauthoryear{{Takahashi} \& {Inutsuka}}{{Takahashi} \&
  {Inutsuka}}{2016}]{ti16}
{Takahashi} S.~Z.,  {Inutsuka} S.-i.,  2016, \mn@doi [\aj]
  {10.3847/0004-6256/152/6/184}, \href
  {http://adsabs.harvard.edu/abs/2016AJ....152..184T} {152, 184}

\bibitem[\protect\citeauthoryear{{Takahashi} \& {Muto}}{{Takahashi} \&
  {Muto}}{2018}]{tm18}
{Takahashi} S.~Z.,  {Muto} T.,  2018, \mn@doi [\apj]
  {10.3847/1538-4357/aadda0}, \href
  {http://ads.nao.ac.jp/abs/2018ApJ...865..102T} {865, 102}

\bibitem[\protect\citeauthoryear{{Tanaka}, {Takeuchi}  \& {Ward}}{{Tanaka}
  et~al.}{2002}]{tan02}
{Tanaka} H.,  {Takeuchi} T.,   {Ward} W.~R.,  2002, \mn@doi [\apj]
  {10.1086/324713}, \href {http://adsabs.harvard.edu/abs/2002ApJ...565.1257T}
  {565, 1257}

\bibitem[\protect\citeauthoryear{{Tomida}, {Okuzumi}  \& {Machida}}{{Tomida}
  et~al.}{2015}]{tom15}
{Tomida} K.,  {Okuzumi} S.,   {Machida} M.~N.,  2015, \mn@doi [\apj]
  {10.1088/0004-637X/801/2/117}, \href
  {http://adsabs.harvard.edu/abs/2015ApJ...801..117T} {801, 117}

\bibitem[\protect\citeauthoryear{{Tominaga}, {Inutsuka}  \&
  {Takahashi}}{{Tominaga} et~al.}{2018}]{tom18}
{Tominaga} R.~T.,  {Inutsuka} S.-i.,   {Takahashi} S.~Z.,  2018, \mn@doi
  [\pasj] {10.1093/pasj/psx143}, \href
  {http://adsabs.harvard.edu/abs/2018PASJ...70....3T} {70, 3}

\bibitem[\protect\citeauthoryear{{Tsukamoto}, {Machida}  \&
  {Inutsuka}}{{Tsukamoto} et~al.}{2013}]{tsu13}
{Tsukamoto} Y.,  {Machida} M.~N.,   {Inutsuka} S.-i.,  2013, \mn@doi [\mnras]
  {10.1093/mnras/stt1684}, \href
  {http://adsabs.harvard.edu/abs/2013MNRAS.436.1667T} {436, 1667}

\bibitem[\protect\citeauthoryear{{Tsukamoto}, {Takahashi}, {Machida}  \&
  {Inutsuka}}{{Tsukamoto} et~al.}{2015a}]{tsu15}
{Tsukamoto} Y.,  {Takahashi} S.~Z.,  {Machida} M.~N.,   {Inutsuka} S.,  2015a,
  \mn@doi [\mnras] {10.1093/mnras/stu2160}, \href
  {http://adsabs.harvard.edu/abs/2015MNRAS.446.1175T} {446, 1175}

\bibitem[\protect\citeauthoryear{{Tsukamoto}, {Iwasaki}, {Okuzumi}, {Machida}
  \& {Inutsuka}}{{Tsukamoto} et~al.}{2015b}]{tsu15b}
{Tsukamoto} Y.,  {Iwasaki} K.,  {Okuzumi} S.,  {Machida} M.~N.,   {Inutsuka}
  S.,  2015b, \mn@doi [\apjl] {10.1088/2041-8205/810/2/L26}, \href
  {http://adsabs.harvard.edu/abs/2015ApJ...810L..26T} {810, L26}

\bibitem[\protect\citeauthoryear{{Tsukamoto}, {Okuzumi}  \&
  {Kataoka}}{{Tsukamoto} et~al.}{2017}]{tsu17}
{Tsukamoto} Y.,  {Okuzumi} S.,   {Kataoka} A.,  2017, \mn@doi [\apj]
  {10.3847/1538-4357/aa6081}, \href
  {http://adsabs.harvard.edu/abs/2017ApJ...838..151T} {838, 151}

\bibitem[\protect\citeauthoryear{{Uribe}, {Klahr}, {Flock}  \&
  {Henning}}{{Uribe} et~al.}{2011}]{uri11}
{Uribe} A.~L.,  {Klahr} H.,  {Flock} M.,   {Henning} T.,  2011, \mn@doi [\apj]
  {10.1088/0004-637X/736/2/85}, \href
  {http://adsabs.harvard.edu/abs/2011ApJ...736...85U} {736, 85}

\bibitem[\protect\citeauthoryear{{Vorobyov}}{{Vorobyov}}{2009}]{vor09}
{Vorobyov} E.~I.,  2009, \mn@doi [\apj] {10.1088/0004-637X/704/1/715}, \href
  {http://adsabs.harvard.edu/abs/2009ApJ...704..715V} {704, 715}

\bibitem[\protect\citeauthoryear{{Vorobyov} \& {Basu}}{{Vorobyov} \&
  {Basu}}{2010}]{vb10}
{Vorobyov} E.~I.,  {Basu} S.,  2010, \mn@doi [\apjl]
  {10.1088/2041-8205/714/1/L133}, \href
  {http://adsabs.harvard.edu/abs/2010ApJ...714L.133V} {714, L133}

\bibitem[\protect\citeauthoryear{{Wada}, {Tanaka}, {Okuzumi}, {Kobayashi},
  {Suyama}, {Kimura}  \& {Yamamoto}}{{Wada} et~al.}{2013}]{wad13}
{Wada} K.,  {Tanaka} H.,  {Okuzumi} S.,  {Kobayashi} H.,  {Suyama} T.,
  {Kimura} H.,   {Yamamoto} T.,  2013, \mn@doi [\aap]
  {10.1051/0004-6361/201322259}, \href
  {http://adsabs.harvard.edu/abs/2013A%26A...559A..62W} {559, A62}

\bibitem[\protect\citeauthoryear{{Ward}}{{Ward}}{1997}]{war97}
{Ward} W.~R.,  1997, \mn@doi [\icarus] {10.1006/icar.1996.5647}, \href
  {http://adsabs.harvard.edu/abs/1997Icar..126..261W} {126, 261}

\bibitem[\protect\citeauthoryear{{Wetherill} \& {Stewart}}{{Wetherill} \&
  {Stewart}}{1989}]{ws89}
{Wetherill} G.~W.,  {Stewart} G.~R.,  1989, \mn@doi [\icarus]
  {10.1016/0019-1035(89)90093-6}, \href
  {http://adsabs.harvard.edu/abs/1989Icar...77..330W} {77, 330}

\bibitem[\protect\citeauthoryear{{Williams} \& {Cieza}}{{Williams} \&
  {Cieza}}{2011}]{wc11}
{Williams} J.~P.,  {Cieza} L.~A.,  2011, \mn@doi [\araa]
  {10.1146/annurev-astro-081710-102548}, \href
  {http://adsabs.harvard.edu/abs/2011ARA%26A..49...67W} {49, 67}

\bibitem[\protect\citeauthoryear{{Wurster}, {Price}  \& {Bate}}{{Wurster}
  et~al.}{2016}]{wur16}
{Wurster} J.,  {Price} D.~J.,   {Bate} M.~R.,  2016, \mn@doi [\mnras]
  {10.1093/mnras/stw013}, \href
  {http://adsabs.harvard.edu/abs/2016MNRAS.457.1037W} {457, 1037}

\bibitem[\protect\citeauthoryear{{Yen}, {Liu}, {Gu}, {Hirano}, {Lee},
  {Puspitaningrum}  \& {Takakuwa}}{{Yen} et~al.}{2016}]{yen16}
{Yen} H.-W.,  {Liu} H.~B.,  {Gu} P.-G.,  {Hirano} N.,  {Lee} C.-F.,
  {Puspitaningrum} E.,   {Takakuwa} S.,  2016, \mn@doi [\apjl]
  {10.3847/2041-8205/820/2/L25}, \href
  {http://adsabs.harvard.edu/abs/2016ApJ...820L..25Y} {820, L25}

\bibitem[\protect\citeauthoryear{{Youdin}}{{Youdin}}{2011}]{you11}
{Youdin} A.~N.,  2011, \mn@doi [\apj] {10.1088/0004-637X/731/2/99}, \href
  {http://adsabs.harvard.edu/abs/2011ApJ...731...99Y} {731, 99}

\bibitem[\protect\citeauthoryear{{Youdin} \& {Goodman}}{{Youdin} \&
  {Goodman}}{2005}]{yg05}
{Youdin} A.~N.,  {Goodman} J.,  2005, \mn@doi [\apj] {10.1086/426895}, \href
  {http://adsabs.harvard.edu/abs/2005ApJ...620..459Y} {620, 459}

\bibitem[\protect\citeauthoryear{{Youdin} \& {Lithwick}}{{Youdin} \&
  {Lithwick}}{2007}]{yl07}
{Youdin} A.~N.,  {Lithwick} Y.,  2007, \mn@doi [\icarus]
  {10.1016/j.icarus.2007.07.012}, \href
  {http://adsabs.harvard.edu/abs/2007Icar..192..588Y} {192, 588}

\makeatother
\end{thebibliography}

%\begin{thebibliography}{99}
%\end{thebibliography}

%%%%%%%%%%%%%%%%%%%%%%%%%%%%%%%%%%%%%%%%%%%%%%%%%%

%%%%%%%%%%%%%%%%% APPENDICES %%%%%%%%%%%%%%%%%%%%%

%\appendix



%%%%%%%%%%%%%%%%%%%%%%%%%%%%%%%%%%%%%%%%%%%%%%%%%%

% Don't change these lines
\bsp	% typesetting comment
\label{lastpage}
\end{document}